\documentclass[final,5p,times,twocolumn,square,numbers]{elsarticle}
\usepackage{amssymb}
\usepackage{amsmath}
\usepackage{graphicx}
\usepackage{balance}
\usepackage{footnote}
\usepackage{multirow} 
\usepackage{float}
\usepackage{algorithm}
\usepackage{algorithmic}
\usepackage{hyperref}
\usepackage{url}
\hypersetup{
  colorlinks=true,
  linkcolor=blue,
  citecolor=blue,
  urlcolor=blue
}

\makeatletter
\def\ps@pprintTitle{%
 \let\@oddhead\@empty
 \let\@evenhead\@empty
 \let\@oddfoot\@empty
 \let\@evenfoot\@empty
}
\makeatother

\begin{document}

\begin{frontmatter}

\title{BEAM-Net: A Deep Learning Framework with Bone Enhancement Attention Mechanism for High Resolution High Frame Rate Ultrasound Beamforming}

\author[a]{Midhila Madhusoodanan\corref{cor1}}
\ead{midhila@ualberta.ca}
\cortext[cor1]{Corresponding author}
\author[b]{Mahesh Raveendranatha Panicker}
\author[c]{Pisharody Harikrishnan Gopalakrishnan}
\author[a]{Abhilash Rakkunedeth Hareendranathan}


\affiliation[a]{organization={Department of Radiology and Diagnostic Imaging, University of Alberta},
            addressline={Edmonton}, 
            country={Canada}} 
\affiliation[b]{organization={Infocomm Technology, Singapore Institute of Technology}, country={Singapore}}
\affiliation[c]{organization={Indian Institute of Technology Palakkad}, 
            country={India}}


\begin{abstract}
Pocket-sized, low-cost point-of-care ultrasound (POCUS) devices are increasingly used in musculoskeletal (MSK) applications for structural examination of bone tissue. However, the image quality in MSK ultrasound is often limited by speckle noise, low resolution, poor contrast, and anisotropic reflections, making bone images difficult to interpret without additional post-processing. Typically, medical ultrasound systems use delay and sum beamforming (DASB) for image reconstruction, which is not specifically optimized for bone structures. To address these limitations, we propose BEAM-Net, a novel end-to-end deep neural network (DNN) that performs high-frame-rate ultrasound beamforming with integrated bone enhancement, using single-plane-wave (SPW) radio frequency (RF) data as input. Our approach embeds a Bone Probability Map (BPM), which acts as an attention mechanism to enforce higher structural similarity around bony regions in the image. The proposed approach is the first of its kind to incorporate bone enhancement directly into ultrasound beamforming using deep learning. BEAM-Net was trained and evaluated on \texorpdfstring{\textit{in-vivo}}{in-vivo} MSK and synthetic RF ultrasound datasets. This paper introduces the Edge Preservation Index (EPI) as a new region-focused metric for evaluating structural fidelity in bone-enhanced ultrasound images. The performance of BEAM-Net was compared with conventional DASB and existing deep learning architectures using the EPI, Contrast Ratio (CR), Signal-to-Noise ratio (SNR), Speckle Similarity Index (SSI), and Structural Similarity Index (SSIM). 
BEAM-Net showed substantial gains over SPW-DASB, achieving 51.4-51\% higher CR and 94.2-73.3\% higher SNR on \texorpdfstring{\textit{in-vivo}}{in-vivo} MSK and synthetic RF datasets. It outperformed multiple steered plane wave DASB (MPW-DASB), with 19.8-24.0\% improvements in CR and SNR on \texorpdfstring{\textit{in-vivo}}{in-vivo} MSK and 2.5-12.8\% improvements on synthetic data. In comparison with existing deep learning models, BEAM-Net improved the SSI by 12.5\%,  SSIM by 22.56\%, and achieved an  EPI of 0.5\%. Although BEAM-Net was only trained on wrist data it generalized well to noisy and unseen elbow datasets as well. Compared to MPW-DASB, BEAM-Net reconstructs high-quality bone-enhanced images from SPW data much faster \texorpdfstring{($< 3ms$)}{(<3 ms)} making it ideal for deployment on POCUS devices for fast and reliable MSK examinations.

\end{abstract}

\end{frontmatter}

\section{Introduction}
\begin{figure}[ht]
    \centering
    \includegraphics[width=1.0\linewidth,trim={1cm 0cm 10cm 0cm}, clip]{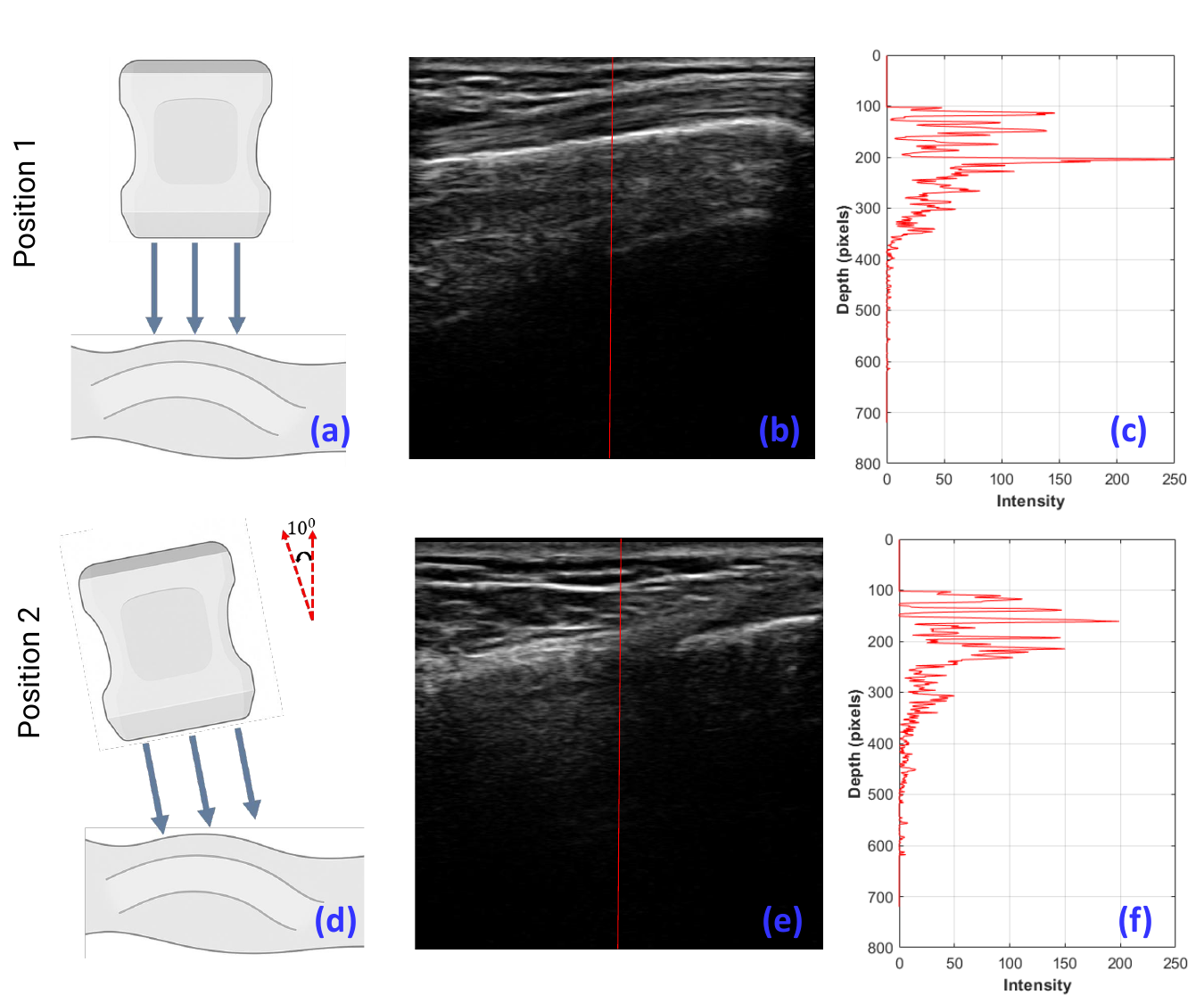}
    \caption{
    Illustration of issues with US probe positioning on the wrist. (a) Position 1: The linear probe is placed perpendicular to the wrist surface, ensuring maximum echo reception. (d) Position 2: The probe is tilted slightly ($<10^\circ$) from the perpendicular orientation, resulting in a variation in the received US intensity.     (b) and (e) show Philips iU22 US images of the radial bone when the probe is placed in Position 1 and Position 2, respectively. (c) and (f) present the corresponding bone scanline profiles for the bone surfaces shown in (b) and (e). The selected scanline is indicated by a red vertical line in each B-mode US image. This setup demonstrates the impact of minor probe alignment on imaging quality.
}
    \label{fig:1}
\end{figure}

Ultrasound (US) imaging is a popular choice for non-invasive examination of musculoskeletal (MSK) structures, soft tissues, organs, and vascular systems, as it avoids ionizing radiation and is significantly less expensive than other imaging modalities such as Magnetic Resonance Imaging (MRI) and Computed Tomography (CT) \cite{Chan2011}. With the advent of portable, low-cost pocket-sized point-of-care ultrasound (POCUS) imaging systems \cite{Kim2017}, US is being increasingly used in emergency care for MSK examination and rheumatology applications \cite{Lee2020}. US imaging typically uses a pulse-echo
approach, where an array of transducer elements emits a US beam (transmit beamforming) and the reflections are recorded by the same transducer array over time to reconstruct the image (receive
beamforming). However, the B-mode images from these devices after reconstruction are often suboptimal for bone visualization, with low quality and imaging artifacts limiting the visibility of clinically relevant structures.
This is due to the specular reflections that occur at the tissue–bone interface due to the size and shape of bones \cite{Brendel2002}. Conventional beamforming approaches assume that the direction of ultrasound beam propagation is perpendicular to the tissue. However, in most MSK applications, the transducer is placed at a specific angle based on the target anatomy. Since bone tissue is a specular reflector, the angle of reflection is governed by Snell's law. As a result, even after delay compensation, the received channel data may remain misaligned, causing partial or complete cancellation of the bone signal during summation. This can cause blurred or thickened bone boundaries with reduced contrast. This issue is particularly evident in the case of curved bone surfaces like the wrist or the shoulder \cite{Jain2004}. 
For instance, Fig.~\ref{fig:1} (b) displays B-mode US images of the radial bone captured with a Philips iU22 probe when the US transducer is aligned perpendicular to the bone surface, as shown in Fig.~\ref{fig:1} (a). Fig.~\ref {fig:1} (e) corresponds to the B-mode images acquired when the transducer is not aligned perpendicularly. Although the images represent the same bone, the features corresponding to the bone surface appear differently due to variations in the scanline profiles. 
Various post-processing techniques, including signal processing methods (\cite{Hacihaliloglu2009}, \cite{Anas2015}, \cite{Singh2020}) and deep learning-based methods (\cite{Hareendranathan2022}, \cite{Wang}) have been proposed to improve the image quality of US bone images. 
\subsection{Pixel Intensity Based Approaches}
Earlier approaches were based on image registration and applied in the context of computer-assisted orthopedic surgery (\cite{Brendel2002}, \cite{Clarke2010}, \cite{Ionescu1999}, \cite{Tran2010}, \cite{Masson2011}). Amin et al. \cite{Amin2003} proposed an image registration method for bone surfaces to enhance surgical navigation by aligning US-derived bone surfaces with preoperative CT or MRI model using the iterative closest point (ICP) algorithm. Their approach used edge detection and gradient-based filtering to identify bone boundaries while mitigating speckle noise and acoustic shadowing artifacts.  Yan et al \cite{Yan2011} used a backward tracing method for bone surface detection in spine US images. Using the bone boundary, the US images are registered to CT for surgical planning.   Kowal et al. \cite{Kowal2007} proposed a two-step automated bone contour detection. First, they identified bone regions in the image and then applied edge detection and thresholding to extract the bone contours and used them for surgical planning. Similarly, Foroughi et al. \cite{Foroughi2007} introduced a two-stage segmentation method based on dynamic programming for the detection of the bone surface using the US. While these approaches are useful for surgical planning, they do not inherently improve the quality of the original B-mode image.

Penney et al. \cite{Penney2006} presented an intensity-based US to CT rigid registration approach on cadaveric data by maximizing normalized mutual information (NMI). Since the approach does not account for acoustic shadowing, their accuracy was low near the bone interface. Penney et al address this by applying contrast adjustments as a pre-processing step to improve bone surface visibility. Along similar lines, Masson-Sibut et al. \cite{Masson2011} proposed a graph search-based automatic bone detection technique for computer-assisted intramedullary nailing. 

\subsection{Local Phase Based Bone Detection Techniques}
Unlike the pixel intensity-based approaches mentioned in the previous section, local phase-based approaches are generally less susceptible to various US imaging artifacts \cite{Oppenheim1981}. Local phase features have been used for enhancement and detection of bone structures \cite{Anas2015}, \cite{Hacihaliloglu2009}, \cite{Hacihaliloglu2011}, \cite{Hacihaliloglu2012}, \cite{Hacihaliloglu2014}, \cite{Hacihaliloglu2015}. Optimizing filter parameters is a crucial aspect of local phase-based image enhancement. Hacihaliloglu et al. \cite{Hacihaliloglu2012}  accomplished this by utilizing information from the image domain. By selecting the optimal filter parameters, they improved localization accuracy and increased the algorithm's resistance to common US imaging artifacts. Anas et al.  \cite{Anas2015} introduced a method to enhance bone visualization in US images for percutaneous scaphoid fracture fixations. Their approach used quadrature band-pass filters to estimate local phase symmetry, with high symmetry indicating bone locations. Additionally, they incorporated shadow information beneath bone surfaces to further enhance bone responses. The extracted bone surfaces were then used to register a statistical wrist model to US volumes. 
Singh et al. \cite{Singh2020} proposed an algorithm for bone-aware image enhancement in MSK US using harmonic analysis and surface impedance models to improve bone surface visibility, especially in challenging areas affected by acoustic shadowing. Their approach improved bone segmentation accuracy by enhancing the boundaries between bone and surrounding tissues. Hareendranathan et al. \cite{Hareendranathan2022} introduced a novel approach aimed at enhancing the quality of pediatric US images. Their technique utilizes domain-aware contrastive learning to address the unique challenges of pediatric hip imaging, such as low contrast and subtle bone structures. This method performs unpaired image translation of low-quality US images into high-quality images guided by a Bone Probability Map (BPM). Although these approaches improve image quality, they require an additional post-processing step in the image processing pipeline, which can be difficult to integrate into POCUS devices due to limited computational resources, memory, power, and the need for real-time performance.
With the increased access to the signal processing pipeline, an alternative approach is to improve the image quality during image reconstruction or beamforming from  RF data.
\subsection{US Imaging}
US images are typically reconstructed through Delay and Sum beamforming (DAS-B) {\cite{PERROT2021106309}} where the US signals received by transducer elements are first delay compensated based on their relative distances to the reflecting target. Once the channel data are aligned, signal strength is enhanced and noise is suppressed through summation. Since US pulses contain alternating positive and negative peaks, precise alignment is essential to prevent destructive interference. If neighboring channels are misaligned, summing opposing peaks can lead to partial cancellation of signals. This cancellation is beneficial in soft tissue imaging, where it helps suppress off-axis interference and clutter. However, it is less desirable in MSK US imaging, where strong bone reflections can also be attenuated, reducing boundary clarity.

An efficient technique to overcome this is through plane wave transmissions by
Transmitting plane waves (PW) at multiple steering angles (MPW), and coherently compounding the DASB data to enhance bone boundary delineation and contrast resolution \cite{Montaldo}. Coherent plane-wave compounding imaging (CPWC) from MPW-RF offers higher image quality than single plane wave (SPW) transmission, which uses a single plane wave to enable high frame rates but compromises image resolution and contrast. However, the increased computational load and reduced frame rate of MPW-DASB limits its utility in the imaging of dynamic structures.
\subsection{Deep Learning Approaches for Beamforming}
To enable more efficient real-time imaging, recent research has turned to deep learning-based acceleration of tissue-adaptive and non-linear beamforming techniques. Nair et al. \cite{Nair2020} introduced a deep learning approach that operates directly on raw IQ channel data. The deep neural network (DNN) is designed to simultaneously learn the geometry-based time-to-space migration (TOF correction) and the subsequent beam summing of channels to generate a beamformed image. The time-to-space migration aspect of this process presents a significant challenge, as it differs fundamentally from standard image-to-image mapping. However, the speckle quality of the reconstructed image was noticeably poorer, which may impact the interpretability of fine structural details.

Khan et al. \cite{khan} utilized deep convolutional neural networks (CNN) to improve channel aggregation and beam-summing processes after TOF correction. Kessler and Eldar \cite{Kessler} proposed a strategy incorporating TOF correction in the Fourier domain, enabling efficient processing by utilizing only a subset of Fourier coefficients from the received channel data. Following this pre-processing step, a deep convolutional network performs beam-summing, converting the TOF-corrected channel data into a single beamformed RF image without aliasing artifacts.
Rather than entirely replacing the beamforming process, Luijten et al. \cite{luijten} propose using a DNN as an artificial agent to compute optimal apodization weights, given the received pre-delayed channel signals at the array. By replacing only the bottleneck component in the minimum variance distortionless response (MVDR) beamformer and further constraining the model to achieve a near-distortionless response during training.
Simson et al. \cite{simson} proposed a DNN method to learn the entire beamforming process from sub-sampled RF channel data.

While these deep learning-based approaches have shown promising results in the general US, none of these are optimized for MSK applications. This could be done by incorporating bone contour detection, which involves identifying the boundaries of bone structures, into the beamforming process. This can help mitigate imaging limitations and improve reconstruction accuracy around bony regions, which is what is presented in this work.

\subsection{Contribuitions}
In this work, the authors propose a new DL-based model that integrates bone enhancement with high-resolution B-mode image reconstruction from SPW-RF data. In this method, we leverage the advantages of CPWC by incorporating deep learning techniques, intending to integrate them into POCUS devices. Specifically, the main contributions of this paper are as follows:

\begin{itemize}
    \item This work proposes a novel deep learning-based attention mechanism, BEAM-Net, for high resolution images from SPW transmission (SPW-RF) to enable high-frame rate ultrasound bone imaging.
    
    \item This is the first deep learning-based approach that enhances bone regions directly from raw RF data.
    
    \item BEAM-Net does not assume any specific transducer orientation or bone shape, hence it is  generalizable to other MSK use cases
    
    \item A BPM is embedded to guide the enhancement process using US-derived features:
    \begin{itemize}
        \item \textit{Local Phase (LP)} to emphasize high-intensity bone structures,
        \item \textit{Feature Symmetry (FS)} to capture asymmetry caused by bone reflections,
        \item \textit{Integrated Backscatter (IBS)} to enhance shadow and reverberation effects.
    \end{itemize}

    \item  Beam-Net was trained and evaluated on \textit{in-vivo} MSK and synthetic RF ultrasound datasets.

    \item The proposed BEAM-Net architecture is compared against conventional DASB, existing deep learning architectures, and established bone enhancement methods used as reference ground truth to comprehensively validate its performance and enhancement accuracy.
    
    \item Robustness of the method is evaluated by testing on noisy, unseen elbow datasets, demonstrating consistent performance under varying signal conditions.
\end{itemize}

\section{Methodology}

\subsection{Conventional DASB (C-DASB)} 
The DAS is the conventional and most popular receive beamforming scheme in US imaging. In C-DASB, a uniform linear array consisting of \(N_L\) elements is considered. The spatial coordinates of the \(i^{\text{th}}\) pixel within the imaging medium are represented by \((x_i, z_i)\), where the transducer array is assumed to be located at a depth of \(z_i = 0\). In C-DASB approach, the backscattered signals received by the transducer elements are first corrected for round-trip propagation delays, then apodized, and finally summed to form the beamformed output~\cite{PERROT2021106309}. While SPW-DASB uses a single unfocused plane wave, MPW-DASB employs multiple steered transmissions and generates the final beamformed signal by coherently compounding the DASB images obtained from these transmissions~\cite{Montaldo}.

The transmit delay \(\delta_{\text{tx}}(j)\) for a pixel \(P = (x_i, z_i)\), corresponding to the \(j^{\text{th}}\) transmission steered at an angle \(\theta_j\) is calculated as:
\begin{equation}
\delta_{\text{tx}}(j) = \frac{z_i \cos{\theta_j} + x_i \sin{\theta_j}}{c}
\end{equation}
The optimal set of MPW transmission angles \(\{\theta_1, \theta_2, \ldots, \theta_T\}\) can be estimated using :

\begin{equation}
\{\theta_1, \theta_2, \ldots, \theta_T\} \approx \frac{n\lambda}{L}, \quad n = -\frac{N_L}{2}, \ldots, \frac{N_L}{2} - 1,
\end{equation}
where \(\lambda\) is the wavelength, \(L\) is the aperture length,  and \(n\) is an integer indexing the angle set.

The receive delay is defined as a function of the Euclidean distance between the \(p^{\text{th}}\) pixel and the \(k^{\text{th}}\) transducer element \(\bigl(k \in [1, N_L]\bigr)\) located at \(\bigl(x_k, 0\bigr)\), and is computed as:

\begin{equation}
\delta_{\text{rx}}(k) = \frac{\sqrt{\, z_p^{2} + \bigl(x_p - x_k\bigr)^{2} }}{c}
\end{equation}

The reflected signal at time instant \(t\) received by the \(k^{\text{th}}\) transducer element, denoted \(s_i(t)\), for the \(j^{\text{th}}\) transmission, must be compensated by the total delay \(\delta_p(i, j)\), given by:

\begin{equation}
\delta_p(k, j) = \delta_{\text{tx}}(j) + \delta_{\text{rx}}(k)
\end{equation}

Final MPW-DASB output \(S_{\text{DAS}}(x_i, z_i)\)  through coherent plane wave compounding for \(T\) transmission, is given as:

\begin{equation}
S_{\text{DAS}}(x_i, z_i) = \sum_{j=1}^{T} \sum_{k=1}^{N_c} A_k(x_i, z_i) \, s_k\left( \delta_p(k, j) \right),
\end{equation}

where \(A_k(x_i, z_i)\) denotes the receive apodization weight for the \(k^{\text{th}}\) transducer element, and \(s_k(\delta_p(i, j))\) is the received signal from the \(k^{\text{th}}\) element during the \(j^{\text{th}}\) PW transmission compensated with the time-of-flight  \(\delta_p(k, j)\).

\subsection{Proposed Bone Enhancement Attention Mechanism Network (BEAM-Net)}
\begin{figure*}[ht]
    \centering
    \includegraphics[width=0.8\linewidth,trim={4.3cm 0cm 5.5cm 0cm}]{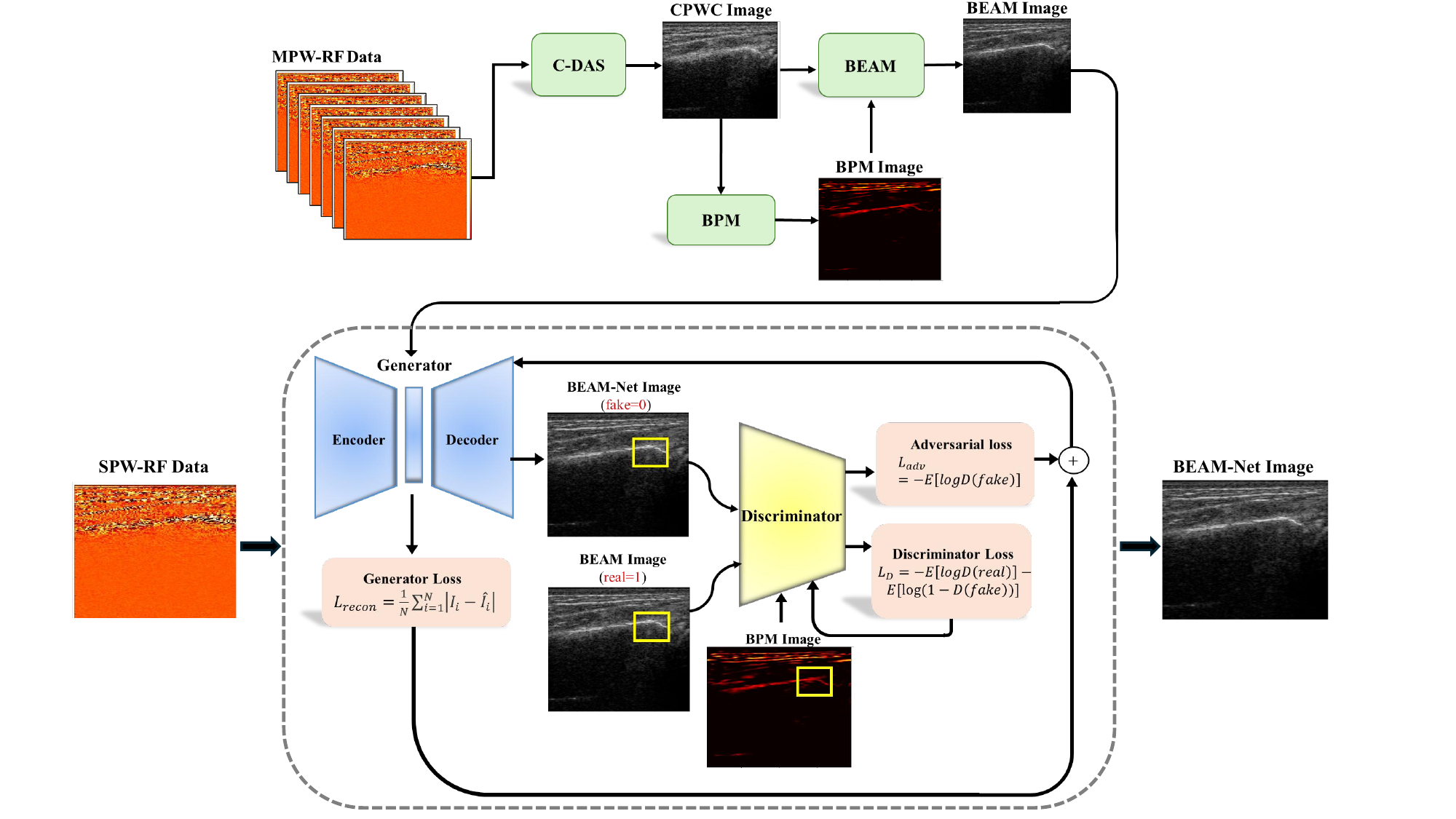}
    \caption{
    Overview of the proposed BEAM-Net processing pipeline. The pipeline illustrates that the processing of MPW-RF data acquired using a US probe is first transformed into a CPWC image, which is further processed using the BEAM (Bone Enhancement Attention Mechanism) method to generate the high-resolution BEAM image. The SPW-RF data and BEAM image are fed into the BEAM-Net model, which enhances the image quality and produces the final high-resolution BEAM-Net image.}
    \label{fig:2}
\end{figure*}
We propose a PatchGAN-based architecture, BEAM-Net (Fig.~\ref{fig:2}), that reconstructs and enhances the high-resolution B-mode US images directly from SPW-RF data, aiming to improve visualization of bone structures. BEAM-Net includes both a generator and a discriminator which are trained simultaneously to optimize both networks. 

The generator uses the U-Net model \cite{Ronneberger2015}, with 3 encoder layers, 3 decoder layers, and skip connections between the encoder and the decoder. The input to the model is RF data of size \( l \times m \times n_{\text{channels}} \), where \( l \) and \( m \) are the number of pixels in the depth and lateral directions, respectively (\( 2194 \times 128 \)), and \( n_{\text{channels}} \) is the number of channels (1 channel).

The discriminator in this architecture follows the PatchGAN approach that evaluates small overlapping patches of the image rather than classifying the entire image as real (ground truth images) or fake (generated images). The discriminator is a deep convolutional neural network (CNN), that processes two-channel input where one channel consists of either a real or generated (fake) image, and the other contains BPM of the image, the conditioning information. This additional channel provides contextual information, which helps the discriminator better distinguish between real and generated images.

The discriminator consists of multiple convolutional layers, each followed by LeakyReLU activation and batch normalization, ensuring stable training. The first layer applies 16 filters of size 4 × 4 with a stride of 2, followed by layers with 64, 128, 256, and 512 filters, progressively increasing feature depth. The final layer uses a single filter with a sigmoid activation, outputting a probability map instead of a single scalar, making it a patch-based discriminator. Unlike traditional discriminators that classify the entire image, PatchGAN outputs a grid where each cell corresponds to a patch.
The discriminator is trained using binary cross-entropy loss (BCE loss), where the real loss encourages the correct classification of real images and the fake loss encourages the correct classification of generated images as fake. The total discriminator loss \( \mathcal{L}_D \) is defined as:

\begin{equation}
\mathcal{L}_D = -\mathbb{E}[\log D({I_{real}})] - \mathbb{E}[\log(1 - D({I_{fake}}))]
\end{equation}

The generator is trained to minimize a weighted sum of two loss components: the adversarial loss \( \mathcal{L}_{\text{adv}} \), which measures the ability of the generator to fool the discriminator, and the reconstruction loss \( \mathcal{L}_{\text{recon}} \), which enforces pixel-level similarity between the generated image and the ground truth. The total generator loss \( \mathcal{L}_G \) is: 
\begin{equation}
\mathcal{L}_G = \mathcal{L}_{\text{adv}} + \lambda \cdot \mathcal{L}_{\text{recon}}
\end{equation}
where 
\begin{equation}
\mathcal{L}_{\text{adv}} = -\mathbb{E}[\log D({I_{fake}})]
\end{equation}
\begin{equation}
\mathcal{L}_{\text{recon}} = \frac{1}{N} \sum_{i=1}^{N} \left| I_{real(i)} - I_{fake(i)} \right|
\end{equation}

Here, \( I_{real} \) denotes the ground truth image and \( I_{fake}\) is the generator’s output, and \(N\) is the total number of pixels over which the loss is computed. We set the weight parameter \( \lambda \) to 100 in all our experiments to emphasize reconstruction accuracy.
The overall algorithm of the  BEAM-Net is presented in Algorithm 1, and notations used are given in Table~\ref{tab:1} 

\begin{algorithm}[ht]
\hrule
\vspace{0.5em}
\caption{Training Algorithm for the Proposed BEAM-Net Model}
\vspace{0.5em}
\textbf{Require:} Training set of input signals $(I_{\text{raw}}, I_{\text{gt}}, I_{\text{bpm}})$, Number of epochs $E$, Batch size $B$ \\
\textbf{Ensure:} Trained generator model $G$
\vspace{0.8em}
\begin{algorithmic}[1]
\STATE \textbf {Initialize generator $G$}
\STATE \textbf{Initialize PatchGAN discriminator $D$}
\STATE Define optimizers for $G$ and $D$ (Adam optimizer, learning rate $1\times10^{-4}$)
\STATE Define loss functions:
\STATE \hspace{1em}\textbf{Discriminator loss (BCE) with BPM input:}

 \hspace{1em} $\mathcal{L}_D = -\mathbb{E}[\log D(I_{\text{gt}}, I_{\text{bpm}})] - \mathbb{E}[\log(1 - D(I_{\text{target}}, I_{\text{bpm}}))]$
\STATE \hspace{1em}\textbf{Generator adversarial loss (BCE):}

 \hspace{1em} $\mathcal{L}_{\text{adv}} = -\mathbb{E}[\log D(I_{\text{target}}, I_{\text{bpm}})]$
\STATE \hspace{1em}\textbf{Generator reconstruction loss (L1):}

\hspace{1em} $\mathcal{L}_{\text{recon}} = \frac{1}{N} \sum_{i=1}^{N} \left| I_{gt(i)} - I_{target(i)} \right|$
\STATE \hspace{1em}\textbf{Total generator loss:}

 \hspace{1em} $\mathcal{L}_G = \mathcal{L}_{\text{adv}} + \lambda \cdot \mathcal{L}_{\text{recon}}$ \\

 \hspace{1em} \textit{where } $\lambda$ \textit{controls the weight of the reconstruction loss (set to 100).}

\FOR{epoch $e = 1$ to $E$}
    \STATE Shuffle training set
    \FOR{each batch}
        \STATE Sample a batch of $(I_{\text{raw}}, I_{\text{gt}}, I_{\text{bpm}})$
        \STATE Generate fake B-mode images: $I_{\text{target}} = G(I_{\text{raw}})$

        \STATE \textbf{Update discriminator}:
        \STATE \hspace{1em} Compute $\mathcal{L}_D$ using $(I_{\text{gt}}, I_{\text{bpm}})$ for real images and $(I_{\text{target}}, I_{\text{bpm}})$ for fake images
        \STATE \hspace{1em} Backpropagate and update $D$
        \STATE \textbf{Update generator}:
        \STATE\hspace{1em} Compute $\mathcal{L}_{\text{adv}}$ and $\mathcal{L}_{\text{recon}}$ 
        \STATE\hspace{1em} Compute total generator loss:
        
         \hspace{1em} $\mathcal{L}_G = \mathcal{L}_{\text{adv}} + \lambda \cdot \mathcal{L}_{\text{recon}}$
        \STATE \hspace{1em} Backpropagate and update $G$
    \ENDFOR
\ENDFOR
\end{algorithmic}
\vspace{0.8em}
\hrule
\vspace{0.5em}
\textbf{Return:} Trained generator $G$
\end{algorithm}
\begin{table}[ht]
\centering
\caption{Notations used in Algorithm 1.}
\label{tab:notations}
\begin{tabular}{ c l }
\hline
\textbf{Notation} & \textbf{Description} \\ \hline
$I_{\text{raw}}$ & RF data \\ \hline
$I_{\text{gt}}$ & BEAM image \\ \hline
$I_{\text{target}}$ & Target image \\ \hline
$I_{\text{bpm}}$ & BPM image \\ \hline
$G$ & Generator model \\ \hline
$D$ & Discriminator model \\ \hline
$\lambda$ & Weight for MSE loss term (typically set to 100) \\ \hline
$E$ & Number of epochs \\ \hline
$B$ & Batch size \\ \hline
$\mathcal{L}_{\text{adv}}$ & Generator adversarial loss\\ \hline
$\mathcal{L}_{\text{recon}}$ & Generator reconstruction loss\\ \hline
$\mathcal{L}_G$ & Total Generator loss \\ \hline
$\mathbb{E}$ & Expectation operator \\ \hline
\end{tabular}
\label{tab:1}
\end{table}

\subsection{Bone Enhancement Attention Mechanism (BEAM)}

In this proposed work, the BEAM image is used as the ground truth for training the network. MPW-RF data acquired from 73 steered PW transmissions (-18° to +18°) were beamformed using DAS and coherently compounded to obtain the CPWC image \(I(x,y)\). Subsequently, BPM \cite{Hacihaliloglu2014} is employed as an image enhancement technique that uses various US features to highlight bone structures in the CPWC image, generating BEAM images (ground truth). 

For computing the BPM, initially integrated backscatter signal \(IBS(x,y)\) \cite{Hacihaliloglu2014} is calculated  by summing the square of the pixel values along each column for the normalized input CPWC image \( I_{\text{norm}}(x,y) \) using:
\begin{equation}
\text{IBS(x,y)} = \sum_{k=1}^{x} I_{\text{norm}}^2 (k,y)
 \label{eq:IBS}
\end{equation}
This integrated measure captures the energy reflected back to the transducer, which is typically higher in bone regions due to the fact that bone tissue acts as a strong reflector. The shadow map is generated by applying a Gaussian weighted averaging scheme along with the image columns. The window of each column is defined by the Gaussian function. 

\begin{equation}
\text{G(i)} = e^{\frac{-i^2}{2\sigma^2}}
 \label{eq:GAUSSIAN}
\end{equation}
where $\sigma$ is a standard deviation parameter that controls the window size. This method smooths the image while preserving the sharp transitions associated with bone boundaries.  

For feature extraction, the Analytic Estimator (AE) is employed, which computes the local phase (LP), feature symmetry (FS), and local energy (LE) \cite{Hacihaliloglu2014}. The LP captures phase variations, which are indicative of the bone surface, while FS quantifies the asymmetry caused by specular reflections at the bone-soft tissue interface. LE measures the signal energy across different frequency bands, providing insights into texture and boundary characteristics.  

The AE method utilizes a multi-scale approach to capture details at different resolutions by applying a series of log-Gabor filters $ G(x, \lambda_0) $ at varying wavelengths. Each filter is tuned to detect features at different scales, where $ \lambda_0 $ is a scaling parameter.
\begin{equation}
G(\omega) = \exp\left(\frac{-\left(\log\left(\frac{|\omega|}{\omega_0}\right)\right)^2}{2 \left(\log(\sigma_0)\right)^2} \right),
\quad \text{where} \quad \omega_0 = \frac{2\pi}{\lambda_0}
 \label{eq:FILTER}
\end{equation}
Symmetric and asymmetric features of the band passed image $I_F (x,y)$ are computed using the Hessian \( H \), Gradient \( \nabla \), and Laplacian \( \nabla^2 \) operations are defined as follows:
\begin{equation}
T_{\text{even}} = [H(I_{\text{F}}(x, y))][H(I_{\text{F}}(x, y))]^T  
 \label{eq:teven}
\end{equation}

\begin{equation}
T_{\text{odd}} = -0.5 \times \left[ \nabla I_{\text{F}}(x, y) \right]  
\left[ \nabla^2 I_{\text{F}}(x, y) \right]^T  
+ \left[ \nabla^2 I_{\text{F}}(x, y) \right]  
\left[ \nabla I_{\text{F}}(x, y) \right]^T
 \label{eq:todd}
\end{equation}
The local phase tensor (\textit{LPT}) \cite{Hacihaliloglu2014} can be calculated by combining the symmetric and asymmetric features with the instantaneous phase $\phi$
\begin{equation}
LPT = \sqrt{T_{\text{even}}^2 + T_{\text{odd}}^2} \times \cos(\phi)
 \label{eq:LPT}
\end{equation}

\begin{figure}[ht]
    \centering
    \includegraphics[width=1.0\linewidth,trim={3cm 11.5cm 5cm 0cm}, clip]{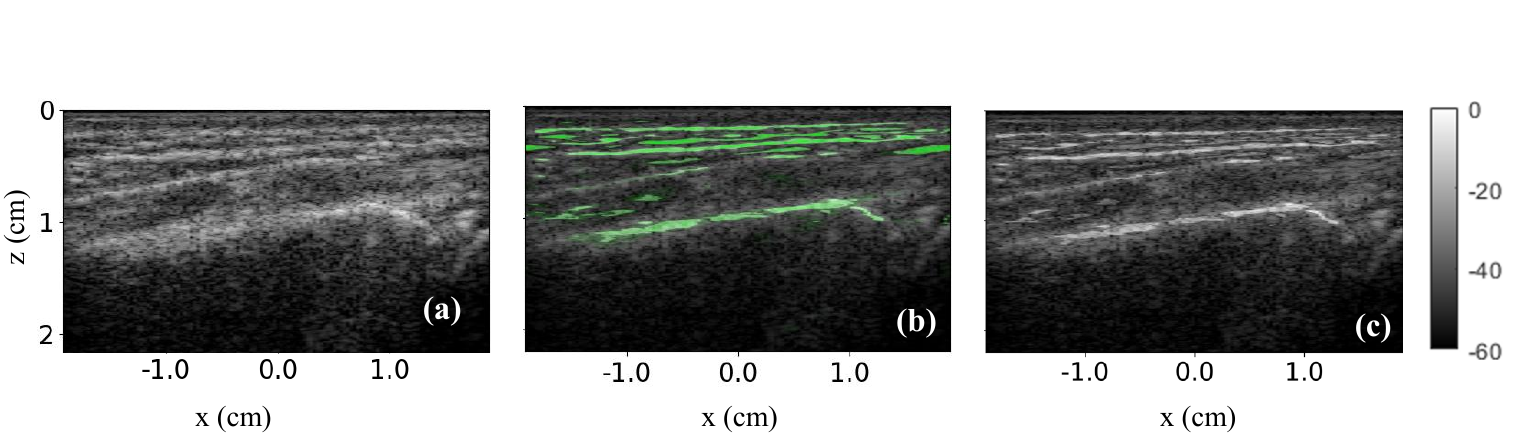}
    \caption{
        (a) CPWC image reconstructed from MPW-RF data, showing the initial visualization of the bone surface.
        (b) BPM, indicated in green, overlaid on the CPWC image to highlight the areas with a high likelihood of bone presence.
        (c) Enhanced image produced by the BEAM method, showcasing improved bone boundary delineation.}
    \label{fig:3}
\end{figure}

By applying the Riesz transform to the \textit{LPT}, as described in ~\cite{Hacihaliloglu2014}, the \textit{LP} and \textit{FS} using the monogenic signals \( M_1 \), \( M_2 \), and \( M_3 \).
\textit{LP} and \textit{FS} are defined as
\begin{equation}
LP(x) = 1 + \arctan \left( \frac{\sqrt{M_2^2 + M_3^2}}{M_1} \right)
 \label{eq:LP}
\end{equation}

\begin{equation}
FS(x) = \frac{\max (\text{T}_{\text{even}} - \text{T}_{\text{odd}} - \tau)}{M_1^2 + M_2^2 + M_3^2}
\label{eq:FS}
\end{equation}
where \textit{M1} is the even component and \textit{M2} and \textit{M3} are the odd
components are given by
\begin{equation}
M_1 = \textit{Re} \left\{ \textit{IFFT}_2 (G(\omega)) \right\}
\label{eq:M_1}
\end{equation}
\begin{equation}
M_2 = \textit{Re} \left\{ \textit{IFFT}_2 (R_z(G(\omega))) \right\}
\label{eqM_2}
\end{equation}
\begin{equation}
M_3 = \textit{Im} \left\{ \textit{IFFT}_2 (R_z(G(\omega))) \right\}
\label{eq:M_3}
\end{equation}
where \(\textit{Re}\), \(\textit{Im}\), \(\textit{IFFT}_2\), and \(R_z\) represent the real component, imaginary component, inverse 2-D FFT, and Riesz transform, respectively,  and \(G(\omega)\) is derived from equation~\ref{eq:FILTER}.
The BPM is then computed by combining these features. Specifically, the \textit{LP} and \textit{FS} are multiplied together, and the resulting product is adjusted using the integrated backscatter energy \((1-IBS(x,y))\) to enhance the contrast between bone and surrounding soft tissue. The final BPM is expressed as equation ~\ref{eq:bpm}
\begin{equation}
BPM(x, y) = \frac{\left( LP(x, y) \times FS(x, y) \times \left[ 1 - IBS(x, y) \right] - \text{min} \right)}{\text{max} - \text{min}}
 \label{eq:bpm}
\end{equation}
Here, min and max values denote the minimum and maximum values of the product term across the image, ensuring the BPM is normalized between 0 and 1.
\(BPM(x,y)\) results in a probability map that highlights bony regions (as shown in Fig.~\ref{fig:3} (b)) more prominently, making it easier for subsequent image analysis. Finally, a weighted image enhancement process, attention mechanism, is applied to improve the contrast. The original US image and the thresholded BPM using Otsu’s method for automatic binarization are combined using the weighted sum:
\begin{equation}
I_{\text{enhanced}}(x, y) = \alpha \cdot I(x, y) + \beta \cdot BPM(x, y) + \gamma
 \label{eq:enhanced}
\end{equation}

Where $\alpha$, $\beta$, and $\gamma$ are the attention weights (AW) parameters controlling the contribution of the original image, BPM, and a bias term, respectively. For the proposed method, the AW parameters are set as $\alpha$ = 0.30, $\beta$ = 0.09, and $\gamma$ = 0.50. This combined approach provides a high-resolution bone-enhanced image, referred to as the  BEAM image, as defined in Equation~\ref {eq:enhanced}. As illustrated in Fig.~\ref{fig:3} (c), demonstrates improved bone delineation, making it more suitable for subsequent segmentation and analysis tasks.

\section{Datasets and Evaluation metrics } 
\subsection{Datasets}
The BEAM-net model was trained and tested on both simulated and experimental datasets. The datasets were generated/acquired employing the L11- 5v linear array transducer of 128 elements with a center frequency of 7.6 MHz and sampling frequency of 31.25 MHz

\subsubsection{In-vivo MSK datasets}
 The RF data are acquired with Verasonics Vantage 128 programmable research US system (Verasonics Inc., Kirkland, WA, USA). This study was conducted following the principles outlined in the Helsinki Declaration of 1975, as revised in 2000. We collected 1900 wrist data from 6 healthy volunteers. The DAS beamformed images are reconstructed from RF data
 using MATLAB 2024b (The MathWorks, Natick, MA, USA).
 
\subsubsection{Synthetic RF datasets from clinical US images}
The SIMUS US simulation package, part of the MATLAB UltraSound Toolbox (MUST) \cite{GARCIA2022106726}, was used to generate 1200 synthetic datasets.
Synthetic RF data was generated from clinical US images acquired from 20 subjects using a Philips IU22 probe at the Stollery Children’s Hospital ED, with ethics approval (Pro00077093) and informed consent from parents. The datasets include the 5 wrist bone regions defined by the wrist protocol: (1) dorsal; (2) proximal dorsal; (3) radial; (4) volar; and (5) proximal volar. These clinical B-mode images served as templates for constructing scatterer distributions, and modeling the acoustic properties of bone and surrounding tissues. 

\subsection{Evaluation Metrics}
To evaluate the performance of the proposed  BEAM-Net model, we use the following metrics: Contrast Ratio (CR), Signal to Noise Ratio (SNR), Speckle Similarity Index (SSI), Edge Preservation Index (EPI), and Structural Similarity Index Measure (SSIM). These metrics are used to assess the quality of bone US images, providing insights into the clarity, sharpness, and overall quality of the reconstructed and enhanced bone US images.

\begin{figure}[ht]
    \centering
    \includegraphics[width=1\linewidth,trim={1.0cm 6.0cm 8.5cm 1.5cm}, clip]{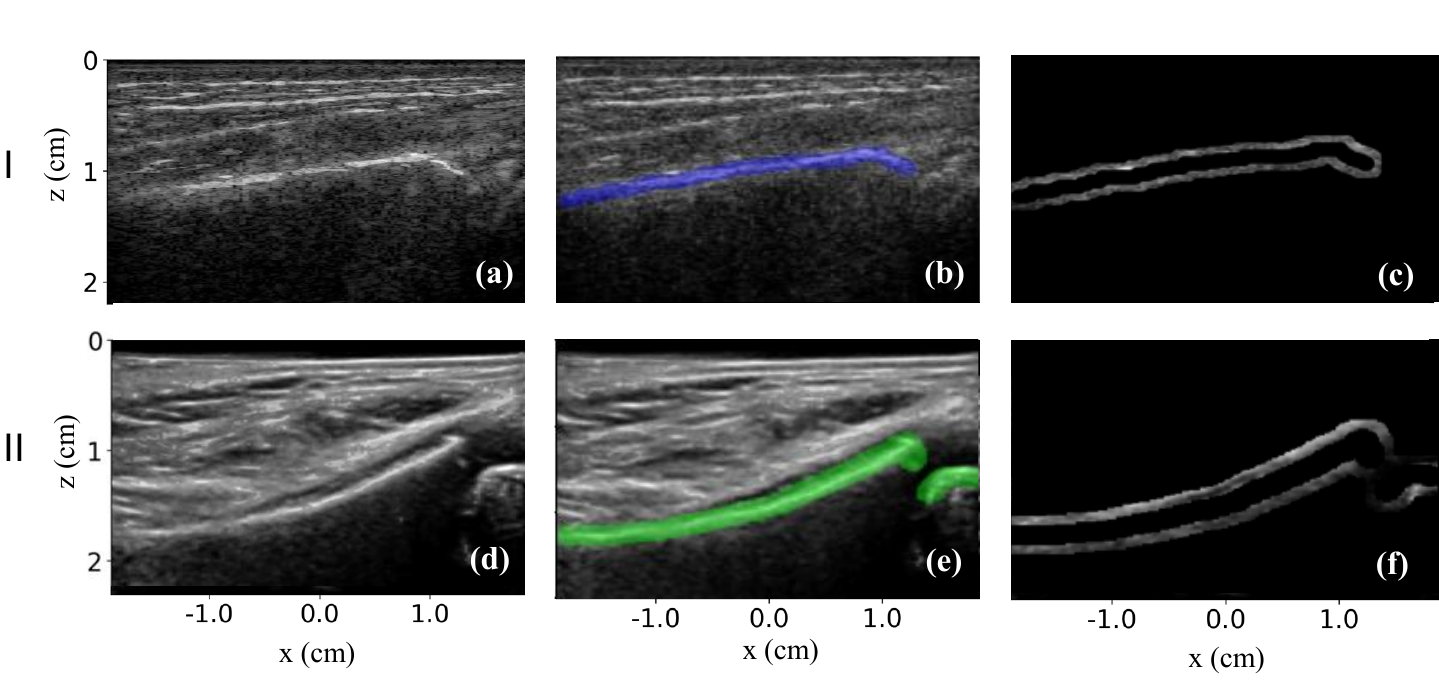}
    \caption{
    Rows I and II show results on the \textit{in-vivo} and synthetic datasets, respectively. From left to right: (a) and (d) display the enhanced US images of the wrist; (b) shows the manually segmented foreground region representing the wrist bone boundary (highlighted in blue), and (e) shows the clinically segmented foreground (highlighted in green); and (c) and (f) depict the background regions obtained by applying morphological dilation to the foreground mask and subtracting the original region to include adjacent non-bone tissue.}
    \label{fig:4}
\end{figure}

The regions of interest (ROIs) for metric computation were defined around the bony structures. Clinically relevant regions in the wrist bone, like epiphysis and metaphysis, were manually segmented for \textit{in-vivo}, as illustrated in Fig.~\ref{fig:4}. For the synthetic RF datasets, images generated from clinical scans were segmented using ground 
truth annotations provided by expert radiologists.
The foreground was defined as the segmented bone region (depicted in blue in Fig.~\ref{fig:4} (b) for \textit{in-vivo} and green in Fig.~\ref{fig:4} (e) for US image from synthetic datasets), while the background was derived by applying morphological dilation to the foreground and subtracting the original region to isolate the adjacent non-bone tissue (as shown in Fig.~\ref{fig:4} (c) and (f)).
\subsubsection{Contrast Ratio (CR)}
CR is used to evaluate the quality of the bone structure by defining an inner and outer region of the bone and using
\begin{equation}
    \text{CR (dB)} = 20 \log_{10} \left( \frac{\mu_{in}}{\mu_{out}} \right)     \label{eq:CNR}
\end{equation}
where \(\mu_{in}\) and \(\mu_{\text{out}}\) represent the mean intensity of the inner and outer regions of the bone, respectively.
\subsubsection{Signal to Noise Ratio(SNR)}
SNR is defined as the ratio of the expected intensity from the bone region to the expected intensity from the background, as given by the following equation, where \(\mu_{in}\) and \(\sigma_{in}\) are the mean and standard deviation of the bone region, and \(\mu_{out}\) and \(\sigma_{out}\) are the mean and standard deviation of the background region, respectively:

\begin{equation}
\mathrm{SNR}(dB) = 10 \log_{10} \left( \frac{\mu_{in}^2 + \sigma_{in}^2}{\mu_{out}^2 + \sigma_{out}^2}\right)
\end{equation}

\subsubsection{Speckle Similarity index (SSI)}
SSI is defined as the sum of the minimum values of the corresponding histogram bins between the ground truth image and the predicted image \cite{Swain1991}. The SSI formula is given by:
\begin{equation}
    SSI(x, y) = \sum_{i=1}^{N} \min(h_{\text{gt}}(i), h_{\text{pred}}(i))
    \label{eq:ssi}
\end{equation}
where \( h_{\text{gt}}(i) \) is the normalized histogram value of the ground truth image at bin \( i \), \( h_{\text{pred}}(i) \) is the normalized histogram value of the predicted image at bin \( i \), \( N \) is the total number of histogram bins.

We adopt the histogram intersection method for measuring similarity, where the similarity is calculated by summing the minimum values of corresponding histogram bins. This method is applied to evaluate the similarity between the speckle patterns in the ground truth and predicted images.
\subsubsection{Structural Similarity Index Measure (SSIM)}
SSIM is used to calculate the structural similarity between ground truth (\textit{X}) and the predicted (\textit{Y}) images. It is defined as:

\begin{equation}
\mathrm{SSIM}(X, Y) = \frac{\bigl(2\mu_X\, \mu_Y + c_1 \bigr)\,\bigl(2\sigma_{XY} + c_2 \bigr)}{\bigl(\mu_X^2 + \mu_Y^2 + c_1 \bigr)\,\bigl(\sigma_X^2 + \sigma_Y^2 + c_2 \bigr)}
\end{equation}

where \(\mu\), \(\sigma\), and \(\sigma^2\) represent the mean, standard deviation, and covariance of the images, respectively. The constants \(c_1\) and \(c_2\) were set to 0.01 to stabilize the division in cases where the denominator is close to zero.

\subsubsection{Proposed Edge Preservation Index (EPI)}
To assess the edge preservation performance of different enhancement methods in bone US imaging, we utilize the EPI, adapted from \cite{Santos}. The EPI quantifies how well the bone edges are retained after processing and is defined as~\ref{eq:EPI}: 
\begin{equation}
    EPI (\%) = \frac{\sum\limits_{i=1}^{M} \sum\limits_{j=1}^{N} D_{gt}(i,j) \cdot D_{pred}(i,j)}{\sqrt{\left( \sum\limits_{i=1}^{M} \sum\limits_{j=1}^{N} D_{gt}(i,j)^2 \right) \cdot \left( \sum\limits_{i=1}^{M} \sum\limits_{j=1}^{N} D_{pred}(i,j)^2 \right)}} \times 100
    \label{eq:EPI}
\end{equation}
where \( D_{gt} = (\hat{I}_{gt} - \mathbb{E}[\hat{I}_{gt}]) \), \( D_{pred} = (\hat{I}_{pred} - \mathbb{E}[\hat{I}_{pred}]) \), and \( \hat{I}_{gt} \), \( \hat{I}_{pred} \) denotes the high-pass filtered version of the ground truth and predicted image, obtained using a standard \( 3 \times 3 \) pixel approximation of the Laplacian operator.

\section{Results}
BEAM-Net was trained using the PyTorch framework on 1000 synthetic RF datasets generated from clinical US wrist images from 7 subjects and 1500 \textit{in-vivo} wrist datasets from 2 volunteers, over 100 epochs with a learning rate of 0.0001 and a batch size of 4. An 80-20 split was used for training and validation. 400 \textit{in-vivo} wrist datasets from 4 volunteers and 200 synthetic clinical datasets from 20 subjects were reserved separately for testing, to evaluate BEAM-Net performance. The Adam optimizer is used to train the model. The training and validation were performed iteratively until convergence, i.e., no decrease in validation loss for five consecutive epochs. The model was trained on the Compute Canada Narval Cluster using an NVIDIA A100 GPU.

In the following section, we first present an ablation study on the evaluation of the impact of the network architecture, enhancement techniques, and parameter settings of BEAM-Net. Subsequently, we performed a functional evaluation and analyzed the robustness of BEAM-Net under varying imaging conditions.

\subsection{Ablation Study}
\begin{figure*}[ht]
    \centering   \includegraphics[width=1.0\linewidth, trim=6cm 0cm 4.5cm 0cm, clip]{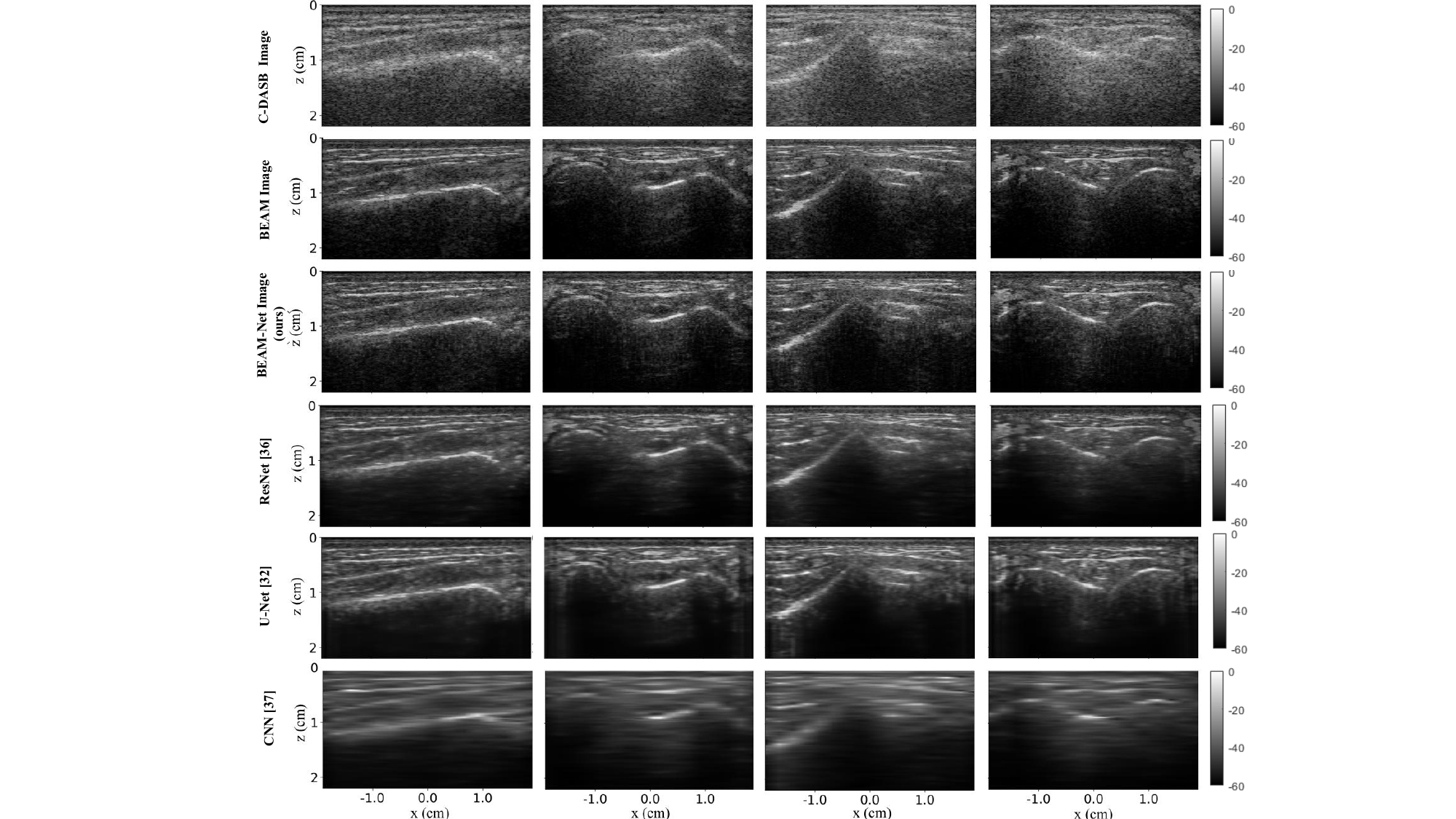}
    \caption{Comparison of deep learning-based enhancement methods on MSK \textit{in-vivo} wrist US data. From the first row to last row: C-DASB image, BEAM image (Ground Truth), BEAM-Net (proposed method), ResNet output, U-Net output,  and CNN output. The figure illustrates the differences in bone boundary clarity and overall image quality achieved by each method.}
    \label{fig:5}
\end{figure*}   
\begin{figure*}[!h]
    \centering   \includegraphics[width=0.8\linewidth, trim=7.5cm 0cm 10cm 0cm, clip]{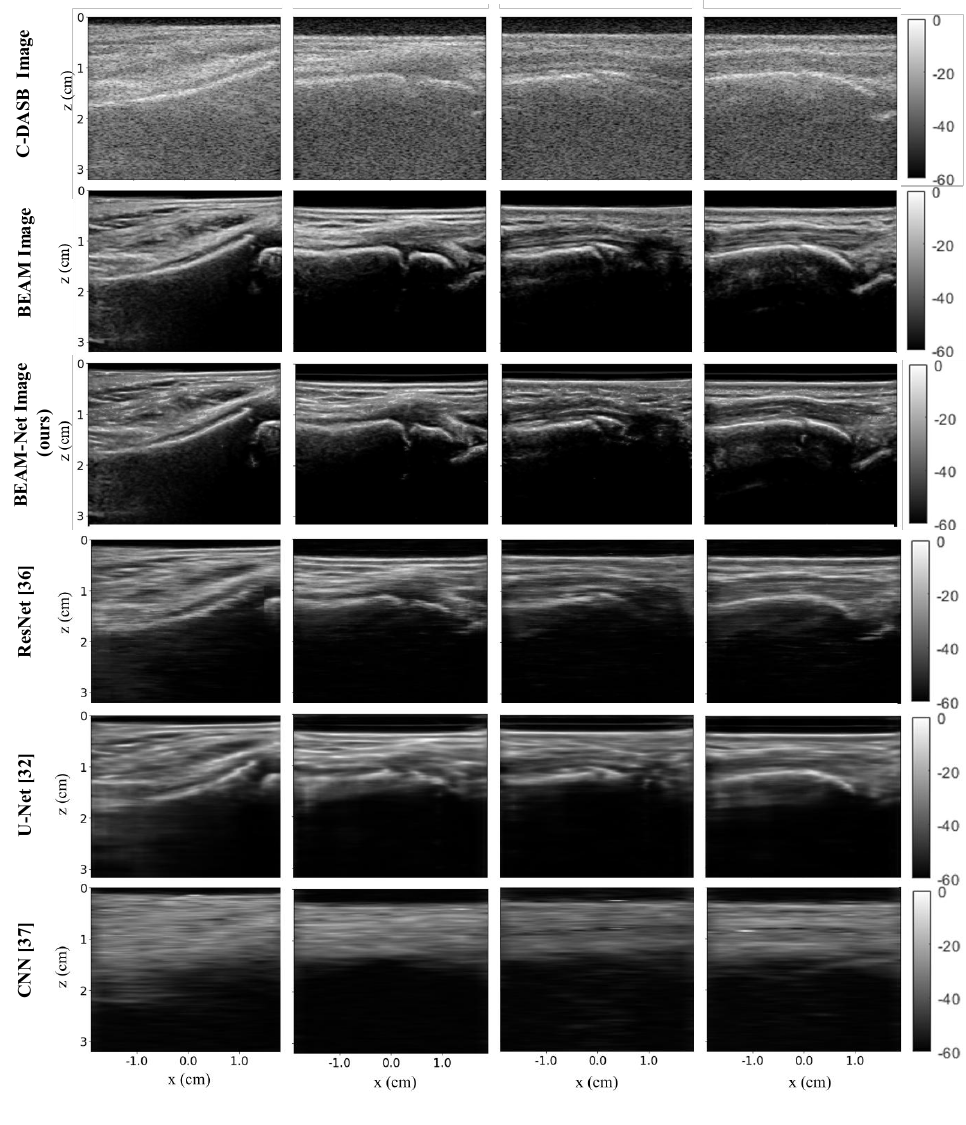}
    \caption{Comparison of deep learning-based enhancement methods on synthetic RF wrist datasets. From the first row to last row: C-DASB image, BEAM image (Ground Truth), BEAM-Net (proposed method),  ResNet output, U-Net output, and CNN output. }
    \label{fig:6}
\end{figure*} 
To investigate the impact of various components on the quality of the reconstructed images in the proposed model, this paper conducts ablation experiments on the network architecture, image enhancement method, and weights of attention parameters. 
\subsubsection{Architecture Ablation}
We compared BEAM-Net with other deep learning methods like  U-Net \cite{Ronneberger2015}, ResNet \cite{7780459}, and CNN \cite{Hyun} using the bone-enhanced, i.e. BEAM image, as ground truth. Results of the quantitative evaluation of different models are presented in  Table~\ref{tab:2} and Table~\ref{tab:3}. The evaluation indices (CR, SNR, SSI, EPI, and SSIM) achieved by our method on the \textit{in-vivo} datasets are 3.49±0.42, 3.46±0.47, 0.94±0.02, 99.92±0.00, and 0.68±0.02, respectively. On the synthetic datasets, these metrics are 7.27±0.87, 7.11±0.95, 0.95±0.01, 99.98±0.01, and 0.95±0.02, respectively. Averaging across experimental and synthetic datasets, BEAM-Net improved the CR, SNR, SSI, SSIM, and EPI metrics by 25.35\%, 23.9\%, 12.5\%, 22.56\%, and 0.5\%, respectively, compared to ResNet.
\begin{table}[ht]
    \centering
    \footnotesize
    \renewcommand{\arraystretch}{1.5}
    \setlength{\tabcolsep}{6pt}
    \begin{tabular}{l
                    cc cc }
        \hline
        \multirow{2}{*}{Method} 
        & \multicolumn{2}{c}{CR (dB)} 
        & \multicolumn{2}{c}{SNR (dB)}  \\
        \cline{2-3} \cline{4-5} 
        & \textit{In-vivo} & Sim 
        & \textit{In-vivo} & Sim 
 \\
        \hline
        C-DASB  
            & 1.71±0.09 & 4.42±0.34
            & 1.67±0.06 & 4.71±0.61  \\
        U-Net 
            & 3.19±0.17 &5.44±0.81 
            & 3.37±0.31 &5.18±0.82
\\
        ResNet 
            & 3.29±0.06* & 5.63±0.87*
            & \textbf{3.50±0.37} & 5.38±0.95* \\
        CNN 
            & 2.70±0.43 & 0.44±0.48
            & 2.90±0.37 & 0.31 ± 0.41 \\
        BEAM-Net
            & \textbf{3.49±0.42} & \textbf{7.27±0.87}
            & {3.46±0.47*} & \textbf{7.11±0.95} \\
        \hline
    \end{tabular}
    \caption{Quantitative comparison (MEAN±STD) of different deep learning methods on \textit{in-vivo} and synthetic RF wrist datasets. The best results are marked with bold text. Asterisks indicate that the difference between our method and the competing method is significant using a paired Student's t-test(*: \((p < 0.05)\)).
    }\label{tab:2}       
\end{table}

\begin{figure}[!ht]
    \centering   \includegraphics[width=1.0\linewidth, trim=6cm 0cm 2cm 0cm, clip]{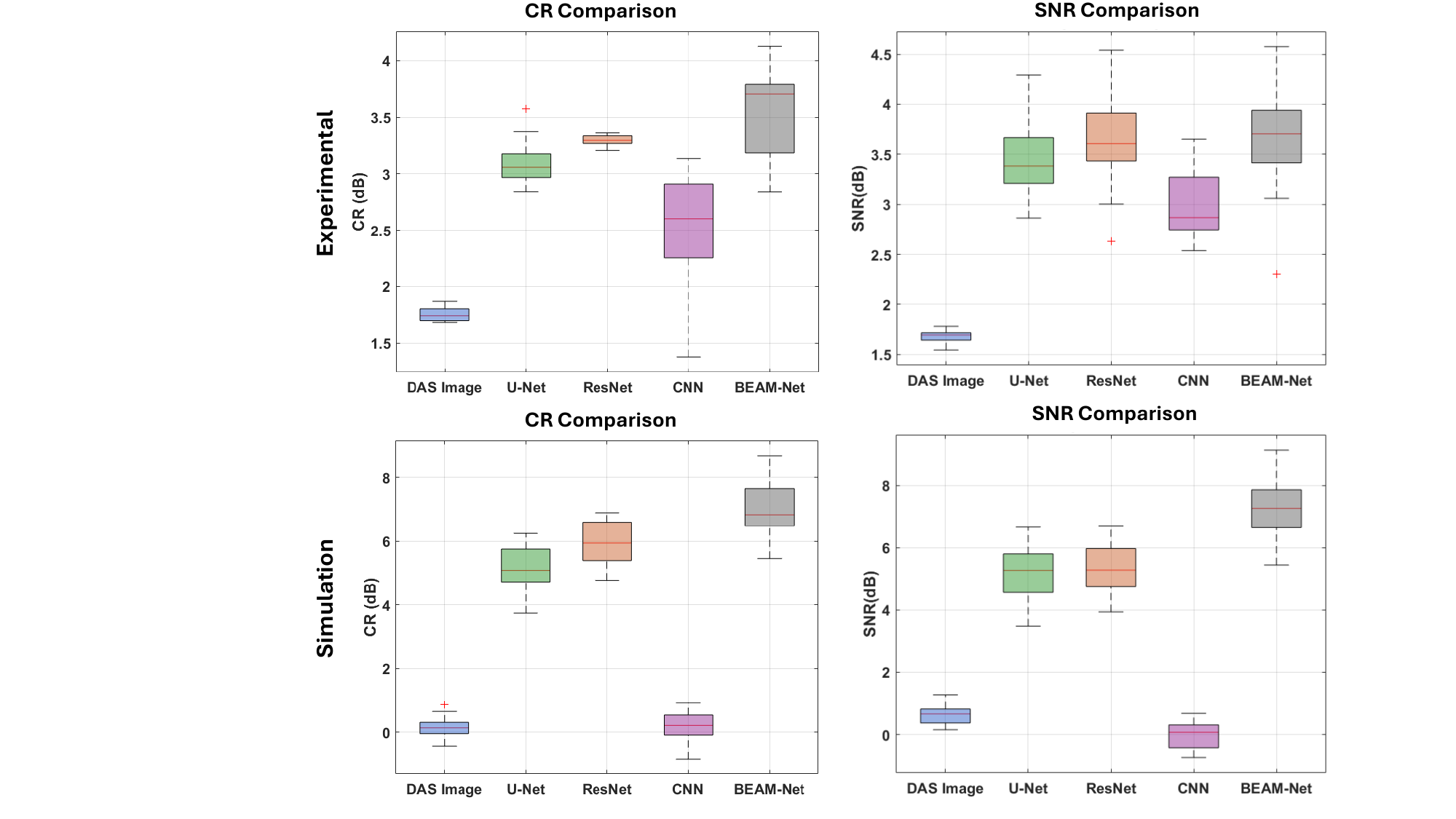}
    \caption{Plots showing the comparison of CR, and SNR for \textit{in-vivo} (Experimental) and synthetic RF wrist (Simulation) US data across different deep learning methods. BEAM-Net demonstrates consistently higher performance across all metrics.}
    \label{fig:7}
\end{figure}

\begin{table*}[!ht]
    \centering
    \footnotesize
    \renewcommand{\arraystretch}{1.5}
    \setlength{\tabcolsep}{2.0pt}
    \begin{tabular}{l
                     cc cc cc}
        \hline
        \multirow{2}{*}{Method}  
        & \multicolumn{2}{c}{SSI} 
        & \multicolumn{2}{c}{EPI (\%)} 
        & \multicolumn{2}{c}{SSIM} \\
        \cline{2-3} \cline{4-5} \cline{6-7}
        & \textit{In-vivo} & Sim 
        & \textit{In-vivo} & Sim 
        &\textit{In-vivo} & Sim \\
        \hline
        U-Net 
            & 0.63±0.03 & 0.89±0.01
            & 96.90±0.10 &99.90±0.01
            & 0.46±0.01 & 0.72±0.02 \\
        ResNet 
            & 0.74±0.05* & 0.94 ±0.03*
            & 99.00±0.07* & 99.91±0.00*
            & 0.60±0.01* & 0.73 ±0.02* \\
        CNN 

            & 0.62±0.01 & 0.58±0.13
            & 94.90±0.20 & 99.87±0.00 
            & 0.39±0.01 & 0.44±0.14 \\
        BEAM-Net
            & \textbf{0.94±0.02} & \textbf{0.95±0.01}
            & \textbf{99.92±0.00} & \textbf{99.98±0.01}
            & \textbf{0.68±0.02} & \textbf{0.95±0.02} \\
        \hline
    \end{tabular}
    \caption{The evaluation metrics (MEAN±STD) of different deep learning methods on \textit{in-vivo} and synthetic RF wrist datasets. The best results are marked with bold text. Asterisks indicate that the difference between our method and the competing method is significant using a paired Student's t-test(*: \((p < 0.05)\)).}\label{tab:3}
    \end{table*}   
\begin{figure}[!ht]
    \centering   \includegraphics[width=1.0\linewidth, trim=0.0cm 1cm 0cm 0cm, clip]{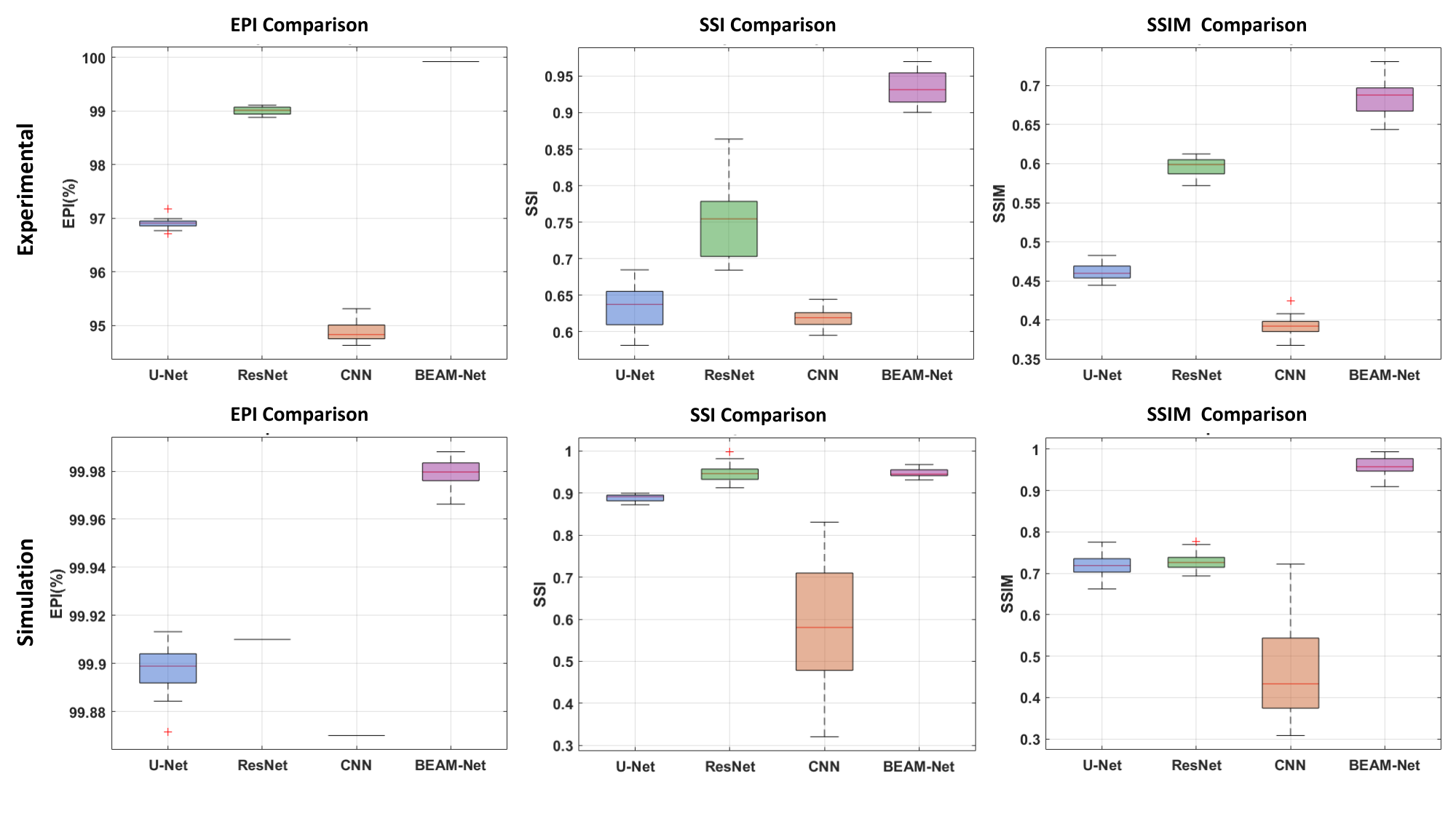}
    \caption{Plots showing the comparison of EPI, SSI, and SSIM metrics for \textit{in-vivo} (Experimental) and synthetic RF wrist (Simulation) US data across different deep learning methods. BEAM-Net demonstrates consistently higher performance across all metrics.}
    \label{fig:8}
\end{figure} 
For each metric, we also conducted a paired Student's t-test comparing our approach against the second-best result. The p-value \((p < 0.05)\) indicates a statistically significant difference between our method and the comparison methods. Fig.~\ref{fig:5} and Fig.~\ref{fig:6} show the visual enhancement results in wrist data from different models on \textit{in-vivo} MSK and synthetic RF datasets. Compared with the reconstruction and enhancement of other deep learning methods, our method achieves superior enhancement results, as illustrated in Fig.~\ref{fig:7} and Fig.~\ref{fig:8}

\subsubsection{Ablation Analysis of Image Enhancement Methods}
\begin{figure}[ht]
    \centering  \includegraphics[width=1.0\linewidth, trim=2.5cm 5cm 9cm 0.5cm, clip]{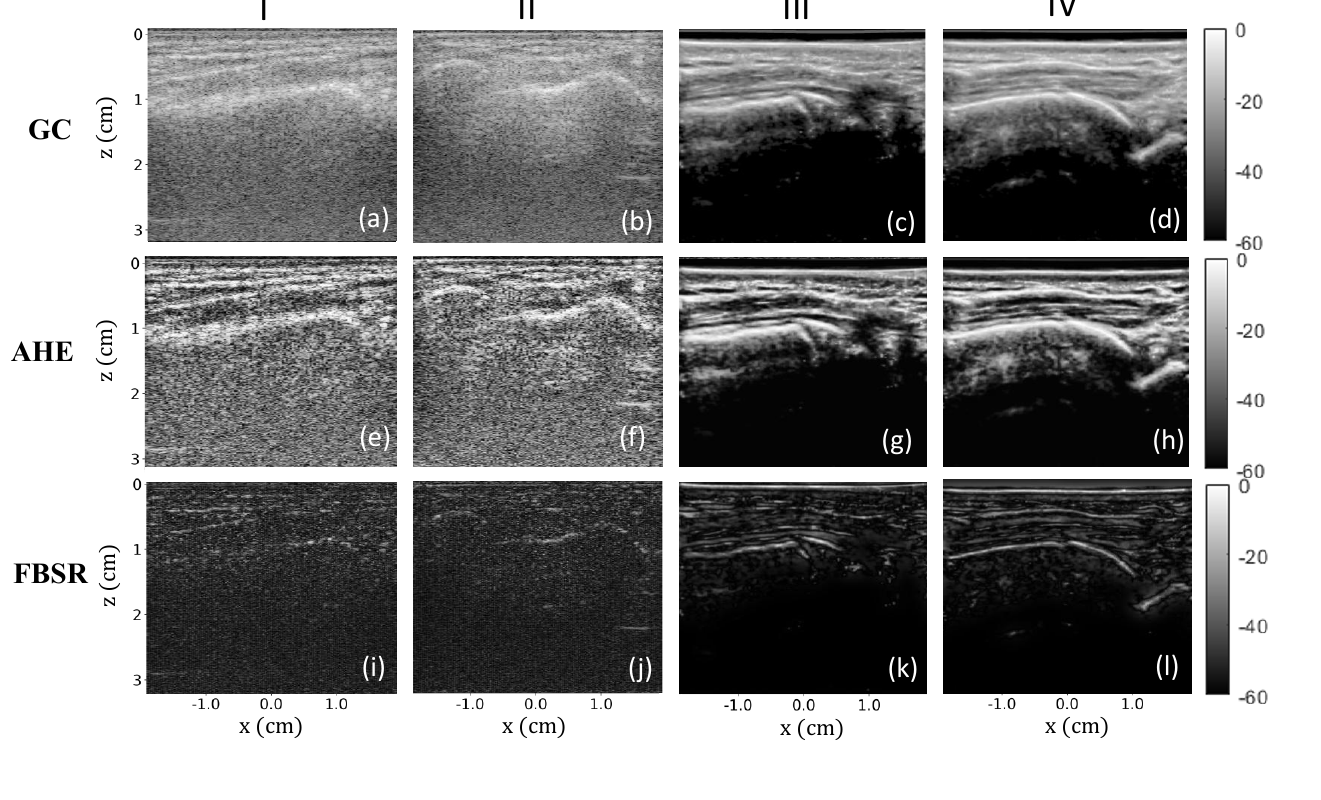}
    \caption{Comparison of different enhancement techniques results by the PatchGAN-based model, applied to wrist US datasets. Columns I, and II show the predicted enhancement results on \textit{in-vivo} datasets. Columns III, and IV present the results on synthetic RF datasets. 
    (a), (b), (c), (d) show results using Gamma Correction (GC); (e), (f), (g), (h) illustrate Adaptive Histogram Equalization (AHE); and (i), (j), (k), (l) display enhancements using Frequency-Based Super-Resolution (FBSR).}
    \label{fig:9}
\end{figure}

We compared the BEAM approach with other image enhancement methods like Gamma Correction (GC), Adaptive Histogram Equalization (AHE), and Frequency-Based Super-Resolution (FBSR) for generating ground truth data and trained using the PatchGAN-based architecture. Table~\ref{tab:4}  presents the evaluation metric of the ablation experiments conducted on the wrist data. The findings indicate that the BEAM technique employed to generate the ground truth image yields superior image quality compared to other techniques. Fig.~\ref{fig:9} visually compares the enhancement results obtained using the PatchGAN-based architecture trained on different ground truths generated by intensity-based techniques.

\begin{table}[!ht]
    \centering
    \footnotesize
    \renewcommand{\arraystretch}{1.5}
    \setlength{\tabcolsep}{4pt}
    \begin{tabular}{l
                    cc cc }
        \hline
        \multirow{2}{*}{Method} 
        & \multicolumn{2}{c}{CR (dB)} 
        & \multicolumn{2}{c}{SNR (dB)}  \\
        \cline{2-3} \cline{4-5} 
        & \textit{In-vivo} & Sim 
        &\textit{In-vivo} & Sim 
 \\
        \hline
        GC 
            & 0.87±0.05  & 3.83±1.63
            & 0.86±0.04 &3.48±1.34 \\
        AHE 
            & 4.24±0.34 & 5.58±2.15
            & 1.72±0.12 &4.69±1.64

\\
        FBSR 
            & 3.16±0.47 &5.00±0.50
            & {2.99± 0.40} & 5.53±0.54
 \\
        BEAM-Net
            & \textbf{3.49±0.42} & \textbf{7.27±0.87}
            & \textbf{3.46±0.47} & \textbf{7.11±0.95} \\
        \hline
    \end{tabular}
    \caption{Evaluation metrics (mean ± standard deviation) of different enhancement methods on the \textit{in-vivo} and wrist dataset predicted by the PatchGAN-based architecture. The best results are highlighted in bold.
    }\label{tab:4}       
\end{table}
\subsubsection{ Attention Weights Parameter Ablation}
To assess the impact of BEAM on ground truth image enhancement, we conduct ablation experiments focusing on the AW parameters. Specifically, we generate multiple sets of ground truth images by varying AW parameter values and use each set to train the benchmark PatchGAN-based architecture. 
The goal is to analyze how each AW component influences the overall network performance. The results of these experiments, summarized in Table~\ref{tab:5}, demonstrate that incorporating appropriate AW parameters enhances the model ability to capture meaningful features, leading to improved image enhancement. Specifically, with AW parameter values $\alpha$ = 0.30, $\beta$ = 0.09, and $\gamma$ = 0.50, BEAM-Net achieved high values in quality performance metrics with a mean CR of 3.49 dB, SNR of 3.46 dB, an SSI of 0.94, and an EPI of 99.92\% on \textit{in-vivo} datasets.

\begin{figure}[!ht]
    \centering
    \includegraphics[width=1.0\linewidth, trim=0.5cm 3.5cm 0.5cm 0cm, clip]{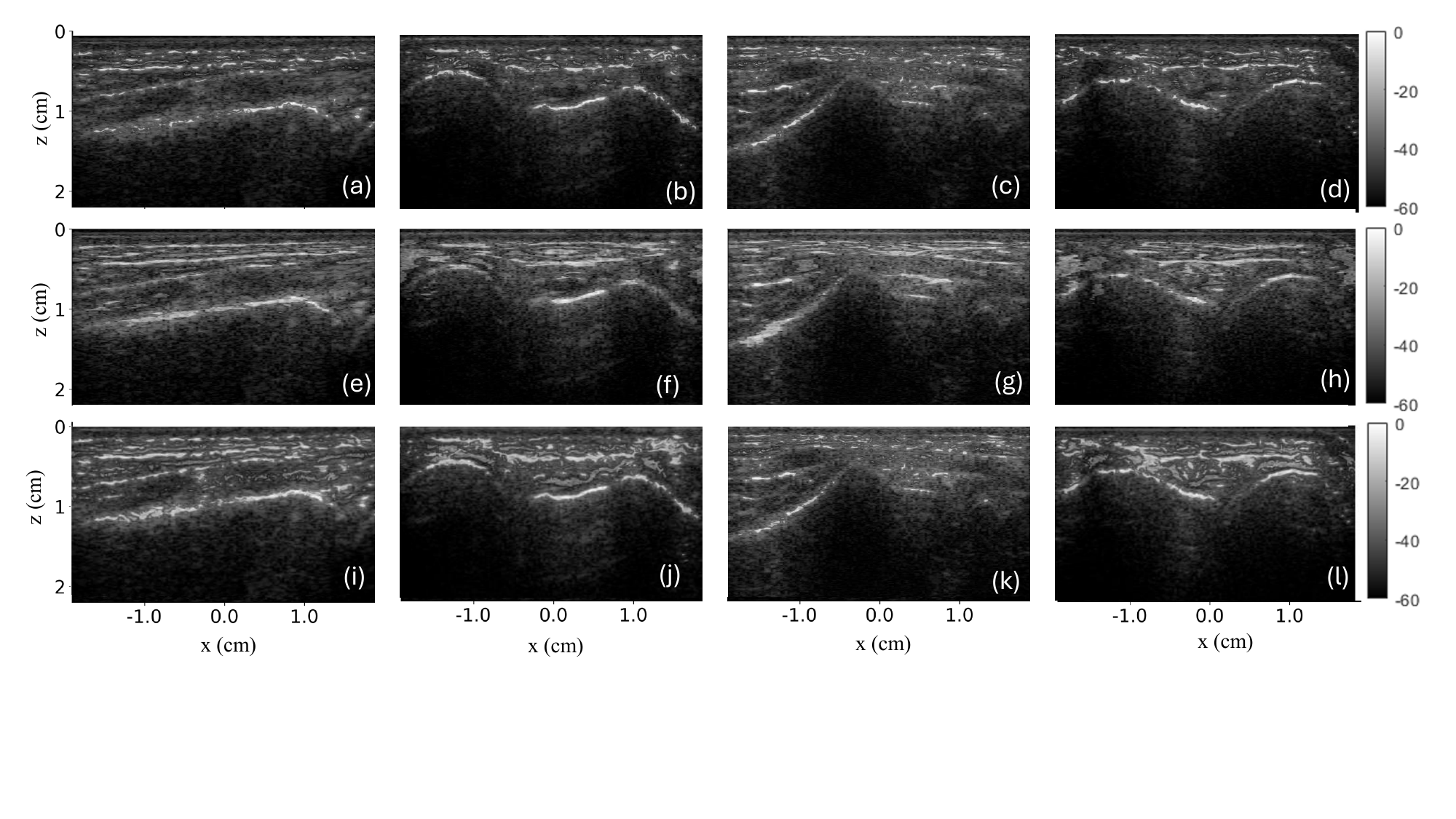}
        \caption{Enhancement results predicted by BEAM-Net with varying attention weight (AW) parameters on \textit{in-vivo} wrist US datasets. The predicted images are based on different sets of AW parameters: (a), (b), (c), (d) show the results predicted with AW parameters $\alpha = 0.20$, $\beta = 0.09$, and $\gamma = 0.5$; (e), (f), (g), (h) display the results predicted with AW parameters $\alpha = 0.30$, $\beta = 0.10$, and $\gamma = 0.5$; (i), (j), (k), (l) present the results predicted with AW parameters $\alpha = 0.40$, $\beta = 0.20$, and $\gamma = 0.5$. }
    \label{fig:10}
\end{figure}

 These findings highlight the effectiveness of our proposed approach in optimizing enhancement techniques for ground truth images.
Fig.~\ref{fig:10} visually compares the enhancement results obtained using different AW configurations, further illustrating the impact of each component on image quality.
\begin{table*}[!ht] 
    \centering
    \footnotesize
    \renewcommand{\arraystretch}{1.5} 
    \setlength{\tabcolsep}{1.2pt} 
    \begin{tabular}{ c c c c c }
        \hline
        Methods & CR (dB) & SNR (dB)  & SSI & EPI (\%)\\
        \hline
         $\alpha$=0.20, $\beta$=0.09, and $\gamma$=0.50   & 2.87±1.03 & 3.15±0.46 & 0.94±0.01 & 99.91±0.00\\
         $\alpha$=0.30, $\beta$=0.10, and $\gamma$=0.50 (Proposed) & \textbf{3.49±0.42} &\textbf{3.46±0.47} & \textbf{0.94±0.02} &\textbf{99.92±0.00}\\
         $\alpha$=0.40, $\beta$=0.20, and $\gamma$=0.50   & 3.10±1.10 & 3.26±0.50 & 0.94± 0.02 & 99.92±0.00 \\
        \hline
    \end{tabular}
    \caption{The evaluation metrics (MEAN±STD) of different parameter settings on BEAM-Net predicted \textit{in-vivo} wrist datasets. The best results are marked with bold text.}
    \label{tab:5}
\end{table*}

\subsection{Fuctional Evaluation}
\begin{figure*}[!ht]
    \centering    \includegraphics[width=1.0\linewidth, trim=1cm 6cm 0.5cm 0cm, clip]{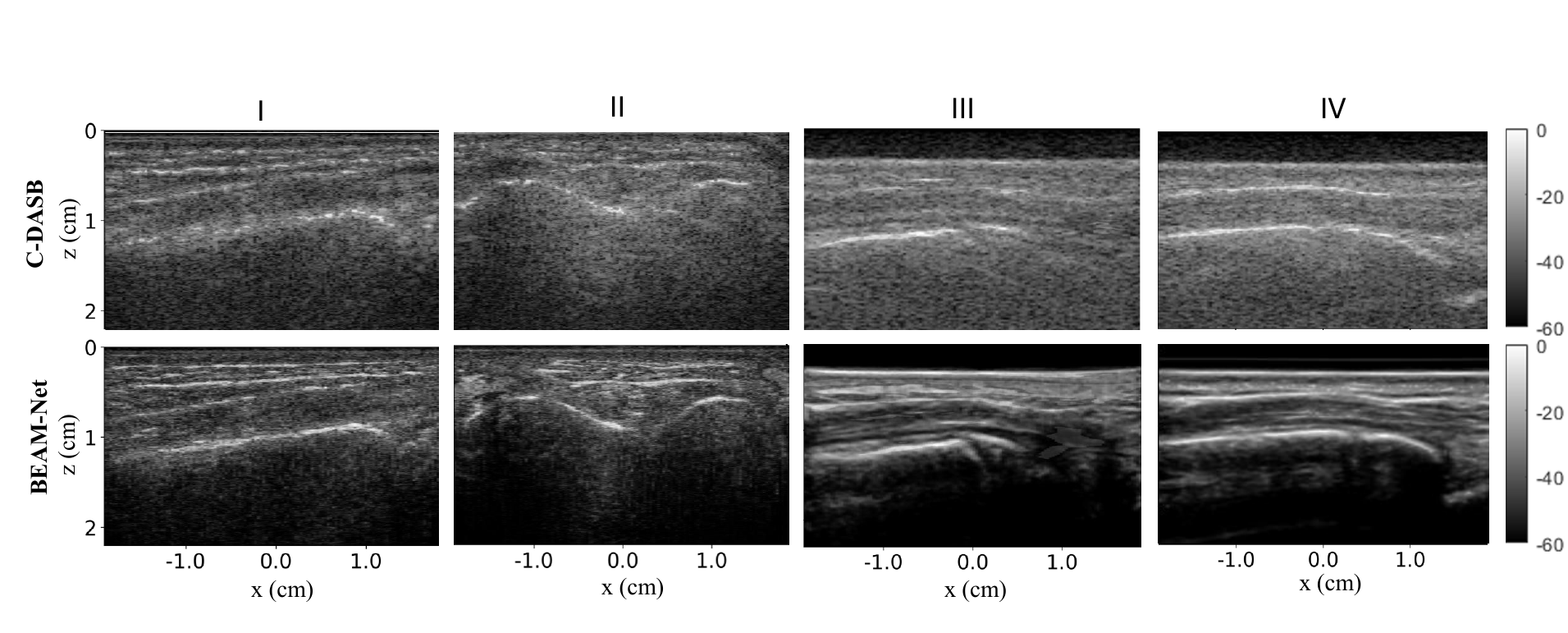}
    \caption{
Comparison of BEAM and BEAM-Net images on SPW-RF data. The first row shows the output images generated by BEAM (Equation~\ref{eq:enhanced}), and the second row shows the corresponding predictions from BEAM-Net. Columns I and II present results on \textit{in-vivo} datasets, while Columns III and IV show results on synthetic RF datasets. The figure highlights the improved bone boundary clarity and overall image quality achieved by the proposed method.
    }
    \label{fig:11}
\end{figure*}

To evaluate the functional contribution of the proposed BEAM-Net architecture, we conducted an ablation study comparing its performance with the direct application of the bone enhancement function (Equation~\ref{eq:enhanced}) on the input data, without the involvement of BEAM-Net. Specifically, the input SPW-RF data is first beamformed using DAS to obtain the B-mode image, and then the BEAM function is applied to generate the bone-enhanced image. Fig.~\ref{fig:11} illustrates this comparison, highlighting the visual and quantitative differences between the two approaches on the same input data. The image produced by directly applying BEAM exhibits significantly poorer quality, with reduced definition of bone structures and lower contrast between bone and surrounding tissues. This degradation is reflected in CR, with performance decreasing by approximately 41\% compared to the BEAM-Net output. This demonstrates that the learned representation within BEAM-Net is capable of reconstructing and enhancing the subtle details of bone structures that the analytical model fails to capture. Furthermore, BEAM-Net effectively learns a complex mapping from low-quality inputs to high-resolution outputs, enabling the network to distinguish between bone and surrounding soft tissues and enhance the bone region with greater precision.
\subsection{Performance Evaluation with MPW-DASB }
\begin{figure}[!ht]
    \centering
    \includegraphics[width=1\linewidth,trim={0.5cm 6cm 7cm 1cm}, clip]{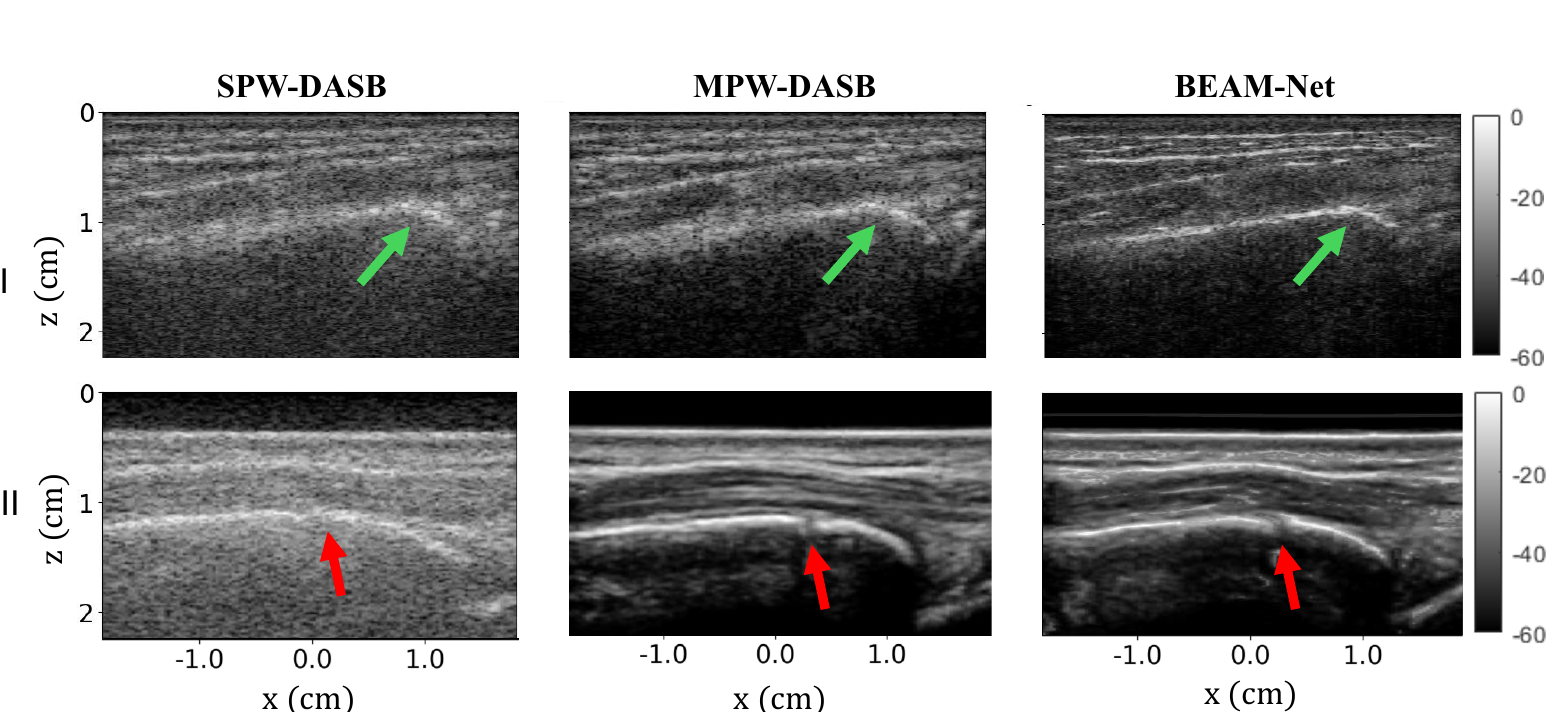}
    \caption{
Comparison of SPW-DASB, MPW-DASB, and BEAM-Net reconstructed images. I shows \textit{in-vivo} data, and II shows synthetic RF data. From left to right: the first column displays the SPW image reconstructed from SPW-RF data, the second column shows the CPWC image reconstructed from MPW-RF data, and the third column presents the enhanced output generated by the proposed BEAM-Net method from SPW-RF data. The BEAM-Net outputs exhibit improved bone boundary delineation, enhanced structural continuity, and higher contrast in both bone and surrounding soft tissue regions.}
    \label{fig:12}
\end{figure}

This study also compares the proposed BEAM-Net model with C-DASB on MPW-RF data, which generates CPWC images. The comparative results are presented in Table~\ref{tab:6}.  As shown in Table~\ref{tab:6}, BEAM-Net achieves a mean CR of 3.29±0.42 dB, which is significantly higher than 2.72±0.19 dB obtained with MPW-DASB. Similarly, BEAM-Net records an SNR of 3.46±0.47 dB, outperforming the MPW-based method, which yields 2.54±0.19 dB. These results indicate that BEAM-Net produces US images with higher contrast and reduced noise levels, even when using SPW-RF data. Furthermore, the reconstruction time required for MPW-DASB is approximately 150 seconds, due to the overhead of coherently compounding the DASB images. In contrast, BEAM-Net reconstructs each image in approximately 3ms using an SPW input, demonstrating significant gains in computational efficiency. A qualitative comparison is illustrated in Fig.~\ref{fig:12}, where BEAM-Net produces clearer bone boundaries compared to MPW-DASB.
\begin{table}[ht] 
    \centering
    \footnotesize
    \renewcommand{\arraystretch}{1.5} 
    \setlength{\tabcolsep}{2pt} 
    \begin{tabular}{l
                    cc cc }
        \hline
        \multirow{2}{*}{Method} 
        & \multicolumn{2}{c}{CR (dB)} 
        & \multicolumn{2}{c}{SNR (dB)}  \\
        \cline{2-3} \cline{4-5} 
        & \textit{In-vivo} & Sim 
        & \textit{In-vivo} & Sim 
 \\
        \hline
        MPW-DASB 
            & 2.72±0.19 & 7.16±0.43 
            & 2.54±0.19 & 6.57±0.14 \\
        BEAM-Net
            & \textbf{3.49±0.42} & \textbf{7.27±0.87}
            &\textbf {3.46±0.47} & \textbf{7.11±0.95} \\
        \hline
    \end{tabular}
    \caption{Quantitative evaluation (MEAN ±STD) of MPW-DASB and BEAM-Net images using various image quality metrics. The best performance for each metric is highlighted in bold.}
    \label{tab:6}
\end{table}   

\subsection{Robustness Analysis}
\begin{figure*}[!h]
    \centering
    \includegraphics[width=1.0\linewidth, trim=0cm 7cm 2cm 0cm, clip]{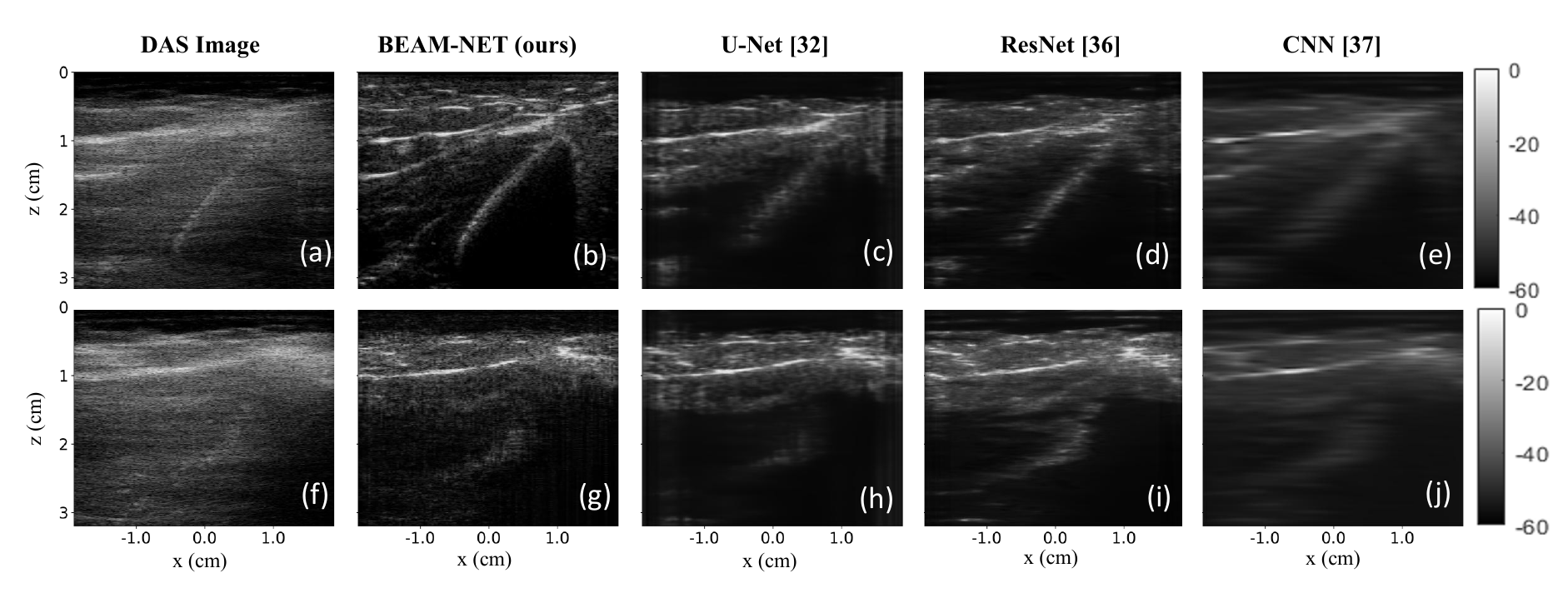}
    \caption{Comparison of deep learning-based enhancement methods on noisy elbow US data. From left to right: (a), (f) C-DAS ; (b), (g) BEAM-Net (proposed method); (c), (h) U-Net output; (d), (i) ResNet output; and (e), (j) CNN output. The figure illustrates the differences in bone boundary clarity and overall image quality achieved by each method when processing noisy input data.}
    \label{fig:13}
\end{figure*}
To further assess the robustness and generalizability of BEAM-Net, we conducted experiments on multiple datasets, focusing on US images of the elbow. The data were acquired using the Verasonics Vantage 128 programmable research US system (Verasonics Inc., Kirkland, WA, USA) with a 128-element linear array transducer (L11-5v) operating at a center frequency of 7.6 MHz and sampling frequency of 31.25 MHz. To evaluate its performance under realistic and challenging conditions, we introduced noise into the test data, simulating low-quality or degraded US acquisitions.
To validate the effectiveness of the enhancement, we compared our method with other deep learning-based approaches, including U-Net \cite{Ronneberger2015}, ResNet \cite{7780459}, and a baseline CNN model \cite{Hyun}. Fig.~\ref{fig:13} presents a visual comparison of the enhancement results and demonstrates that our method can reconstruct and enhance bone structures that are not visible using C-DASB, especially in the presence of noise. 
As shown in Table~\ref{tab:7}, our method consistently outperforms the other models across various quantitative evaluation metrics, achieving a CR of 2.25, SNR OF 1.96, an EPI of 99.86\%, an SSI of 0.80, and an SSIM of 0.40±0.04.

\begin{table*}[ht] 
    \centering
    \footnotesize
    \renewcommand{\arraystretch}{1.5}
    \setlength{\tabcolsep}{5pt} 
    \begin{tabular}{c c c c c c }
        \hline
        Methods & DAS Image  & U-Net & ResNet & CNN& BEAM-Net \\
        &  &  &  & & (Proposed)\\ 
        \hline
        CR (dB) &1.39±0.18  &2.09±0.53  &1.79±0.23  & 1.31±0.40 &\textbf{2.25±0.40} \\
        SNR (dB) & 1.06±0.24  & 1.50±0.30 & \textbf{2.25±0.10}  & 0.83±0.10 &{1.96±0.19 } \\
        SSI & -  &0.65±0.02  &0.69±0.02
  &0.58±0.01& \textbf{0.80±0.01} \\
        EPI (\%) & -  & 90.86±0.01 & 98.86±0.01  & 89.84±0.02& \textbf{99.86±0.01}  \\
        SSIM & -  & 0.35±0.01  &0.37±0.01 &0.34±0.01& \textbf{0.40±0.04} \\
        \hline
    \end{tabular}
    \caption{The evaluation metrics (MEAN±STD) of different deep learning methods on noisy elbow datasets. The best results are marked with bold text.}
    \label{tab:7}
\end{table*}

\section{Discussion}
In this study, we proposed BEAM-Net, a novel deep learning-based attention mechanism for beamforming that enhances bone imaging in US. The objective of BEAM-Net is to generate high-resolution, bone-enhanced images from low-quality raw channel input data. We use a PatchGAN discriminator, which improves the generator ability to produce fine textures and local details, ensuring both global coherence and local consistency in the generated images. This architecture also enhances training stability and reduces the risk of model collapse. To evaluate its effectiveness, the proposed model was quantitatively assessed. The details of this analysis are discussed below.

\subsection{Comparison With Other Architectures}
BEAM-Net images showed statistically significant improvement \((p<0.05)\) in terms of CR, SSI, and EPI when compared to deep learning methods like  U-Net \cite{Ronneberger2015}, ResNet \cite{7780459}, and CNN \cite{Hyun}.  Unlike other DL models, BEAM-Net can preserve natural speckle texture in soft tissue regions while improving bone contrast. Conventional DL models, such as U-Net and CNN, either overly smoothed the image, losing fine structural details, or retained excessive speckle noise that affected image readability. In contrast, BEAM-Net provided sharper bone edges and clearer contrast transitions without introducing artifacts, particularly around anatomically complex areas. Based on the results shown in Fig.~\ref{fig:5} and Fig.~\ref{fig:6}, we can see that the proposed method achieves noticeably better image quality, with enhanced contrast and sharper structural details that are necessary for identifying bone tissue in clinical applications like fracture detection.

\subsection{Comparison with C-DASB}

 \begin{figure*}[ht]
    \centering
    \includegraphics[width=1.0\linewidth,trim={0cm 9cm 1cm 0cm }, clip]{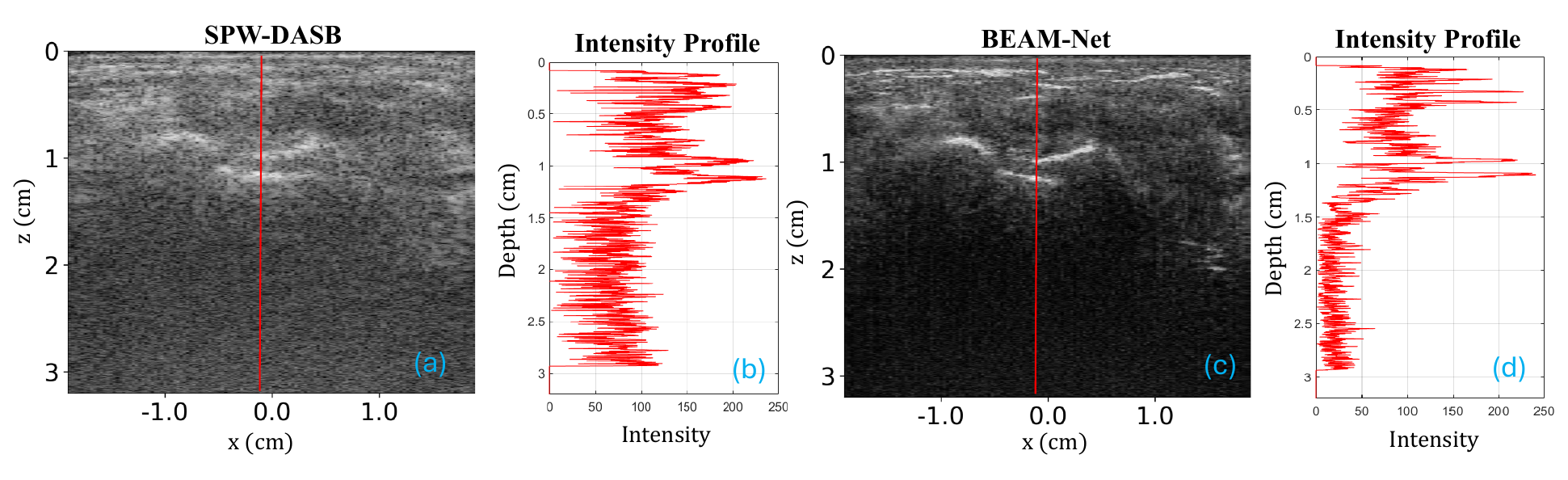}
    \caption{(a) C-DASB image from SPW-RF data and (c) BEAM-Net enhanced image obtained from the same SPW-RF data. Both images depict the same anatomical region, with a red vertical line indicating the selected scanline for bone surface analysis. (b) and (d) shows the corresponding intensity profiles along this scanline, extracted from the DAS and BEAM-Net images, respectively. The BEAM-Net profile demonstrates sharper, more distinct peaks corresponding to bone boundaries, indicating improved structural clarity and edge definition compared to the conventional approach.}
    \label{fig:14}
\end{figure*}

BEAM-Net introduces a novel approach to US beamforming directly from SPW-RF data, eliminating the need for compounded acquisitions as required in MPW-DASB. Despite relying on simplified input, BEAM-Net surpasses MPW-DASB in both qualitative and quantitative performance. In addition to standard improvements in contrast and sharpness, BEAM-Net enhances anatomical details that are often lost or blurred in C-DASB images. As shown in Fig.~\ref{fig:14}, the model reveals subtle features such as fine bone edges and soft tissue interfaces with greater clarity. Quantitative metrics further demonstrate that BEAM-Net improves contrast and reduces noise artifacts. Moreover, BEAM-Net achieves these enhancements with significantly lower computational complexity and faster processing, making it a practical solution for real-time clinical US imaging. By learning spatial coherence patterns typically only accessible through MPW acquisition, BEAM-Net effectively simulates the benefits of MPW-DASB with just SPW-RF data input representing a significant advancement in deep-learning-based beamforming.

\subsection{Generalization and Robustness}
We evaluated the performance of BEAM-Net on noisy test data, generated synthetically to simulate sub-optimal image acquisition conditions. In particular, the test set consisted of RF data from the elbow scans, representing unseen anatomical data, since the model was trained exclusively on wrist datasets. As illustrated in Fig.~\ref{fig:13}, BEAM-Net was able to generate high-quality images even when presented with noisy data. The results summarized in Table~\ref{tab:7} indicate that BEAM-Net not only enhances bone visibility in noisy conditions but also preserves critical anatomical boundaries better than existing approaches. 
  
\subsection{Computational Requirements}

BEAM-Net is also computationally efficient when compared to MPW beamforming, which involves reconstructing multiple steered images followed by coherent compounding. This process is computationally intense and takes approximately 150.16 seconds per image on average, based on our experiments. In contrast, BEAM-Net produces enhanced outputs in under 0.0030 seconds for a depth of 5cm. This drastic reduction in computational load makes BEAM-Net far more practical for real-time clinical applications that have to be performed at a high framerate.

\subsection{Limitations and Future work}
A key limitation of this study is the limited access to RF data, as most  POCUS manufacturers do not provide access to this information. This restricts the broader applicability and validation of the proposed method across different US systems. Additionally, the current study was conducted on a relatively small dataset, which may limit the generalizability of the findings. To address these issues, a clinical study is planned to evaluate the proposed approach on a larger cohort. Furthermore, since BEAM-Net is agnostic to the target anatomy, we intend to extend its application to other MSK US scenarios, such as hip and shoulder imaging, to assess its broader utility and robustness.
\section{Conclusion}
To better address the challenges of bone US image reconstruction and enhancement, we proposed an end-to-end deep learning model, BEAM-Net, inspired by a PatchGAN-based architecture for automatic reconstruction and enhancement of bone US images from SPW-RF data. The proposed model was trained and tested on \textit{in-vivo} and synthetic RF wrist datasets. The image quality of the predicted images was validated with  CR, SNR, SSI, EPI, and SSIM. As evident from the evaluation results, the proposed neural network model, BEAM-Net, showed significant improvement in image quality. Extensive experiments (comparative experiments and robustness analysis) with several deep learning methods demonstrate that our method has better performance on bone image enhancement. This approach could be potentially integrated into POCUS systems to produce high-quality, high-resolution US bone images. 

\section*{Acknowledgment}
The authors thank Dr. Jacob L. Jaremko, Clinician-Scientist and Professor in the Department of Radiology, University of Alberta, and Canada CIFAR AI Chair for insightful discussions on MSK ultrasound. The authors also acknowledge Alberta Innovates AICE Concepts and IC-IMPACTS for funding, and the Indian Institute of Technology Palakkad, India, for their support in data acquisition. We further acknowledge Compute Canada for providing computational resources, including high-performance graphical processing units (GPUs), for training and testing the models. The authors thank Dr. Gayathri Malamal (Research Scientist, FUJIFILM Visualsconics Inc.) for technical discussions and Steel McDonald (Medical Student, University of Alberta) for assistance with data acquisition.

\bibliographystyle{elsarticle-num-names}
\bibliography{elsarticle-template-num-names}


\end{document}